\documentclass[a4paper, 12pt, times, preprint]{elsarticle}

\usepackage{tgtermes}
\usepackage[a4paper,top=2cm,bottom=2cm,left=2cm,right=2cm,marginparwidth=2cm]{geometry}
\usepackage[justification=raggedright,singlelinecheck=false]{caption}
\usepackage{hhline}
\usepackage{float}
\usepackage{multirow}
\usepackage{graphicx}
\usepackage{booktabs}
\usepackage{tabularx}
\usepackage{xfrac}
\usepackage{wrapfig}
\usepackage{mwe}
\usepackage{adjustbox}
\usepackage{caption}
\usepackage{subcaption}
\usepackage{makecell}
\usepackage{rotating}
\usepackage[permil]{overpic}
\usepackage{pict2e}
\usepackage{amsmath,amssymb,array}
\usepackage{xcolor,soul}
\usepackage[most]{tcolorbox}
\usepackage{calrsfs}
\usepackage{longtable}
\usepackage{ragged2e}
\usepackage{siunitx}
\usepackage{tikz}
\usepackage{algorithm}
\usepackage{algpseudocode}
\usepackage{listings}
\usepackage{fancyvrb}
\usepackage[version=3]{mhchem}
\usepackage{enumitem}
\usepackage{caption}
\usepackage{capt-of}
\usepackage{inconsolata} 
\usepackage{microtype}
\usepackage{varwidth}
\usepackage{lineno}

\usepackage{hyperref}  

\hypersetup{
    colorlinks,
    breaklinks,
    urlcolor=Maroon,
    linkcolor=Maroon
}

\newcommand{\etal}{\textit{et al}.}

\definecolor{scalebgcolor}{rgb}{0.08,0.52,0.80}
\definecolor{codegreen}{rgb}{0,0.6,0}
\definecolor{codegray}{rgb}{0.5,0.5,0.5}
\definecolor{codepurple}{rgb}{0.58,0,0.82}
\definecolor{backcolour}{rgb}{0.95,0.95,0.92}

\def\correction#1{%
    \abovedisplayshortskip=#1\baselineskip\relax
    \belowdisplayshortskip=#1\baselineskip\relax
    \abovedisplayskip=#1\baselineskip\relax
    \belowdisplayskip=#1\baselineskip\relax
}

\fboxsep=\tabcolsep\relax
\fboxrule=\arrayrulewidth\relax

\newcolumntype{A}[2]{%
    >{\minipage{\dimexpr#1\linewidth-2\tabcolsep-#2\arrayrulewidth\relax}\vspace\tabcolsep}%
    c<{\vspace\tabcolsep\endminipage}
}

\lstdefinestyle{mystyle}{
    backgroundcolor=\color{backcolour},   
    commentstyle=\color{codegreen},
    keywordstyle=\color{magenta},
    numberstyle=\tiny\color{codegray},
    stringstyle=\color{codepurple},
    basicstyle=\ttfamily\footnotesize,
    breakatwhitespace=false,         
    breaklines=true,                 
    captionpos=b,                    
    keepspaces=true,                 
    numbers=left,                    
    numbersep=5pt,                  
    showspaces=false,                
    showstringspaces=false,
    showtabs=false,                  
    tabsize=2
}
\biboptions{numbers,sort&compress}

\journal{Digital Discovery}

\begin{document}
\begin{frontmatter}

    \title{Agentic Resource Allocation for Batch Multi-Objective Bayesian Optimization in Autonomous Materials Discovery}

    \author[1]{Robert Robinson\corref{corr1}}
    \ead{trippr5118@tamu.edu}
    \cortext[corr1]{Corresponding author}
    
    \author[2]{Shakti Prasad Padhy}
    \author[1]{Sushant Sinha}
    
    \author[1]{Sk Md Ahnaf Akif Alvi}
    \author[2]{Juan Florez Coronel}
    \author[1]{Brent Vela}
    \author[1]{Trevor Hastings}
    
    \author[2]{Douglas Allaire}
    \author[1,2,3]{Raymundo Arr\'oyave}

    \affiliation[1]{organization={Department of Materials Science and Engineering, Texas A\&M University},
                city={College Station},
                postcode={77843}, 
                state={TX},
                country={USA}}

    \affiliation[2]{organization={J. Mike Walker '66 Department of Mechanical Engineering, Texas A\&M University},
                    city={College Station},
                    postcode={77843}, 
                    state={TX},
                    country={USA}}
    
    \affiliation[3]{organization={Wm Michael Barnes '64 Department of Industrial and Systems Engineering},
                    city={College Station},
                    postcode={77843}, 
                    state={TX},
                    country={USA}}

    \begin{abstract}
    The discovery and development of advanced materials is a challenging process constrained by the high time and monetary costs associated with the synthesis, processing, and characterization. Experimentation and high-fidelity simulations are both expensive and time-intensive, and the underlying design spaces can be enormous, often with multiple competing objectives. Bayesian optimization (BO) provides a principled approach for efficiently navigating such spaces, but most workflows rely on fixed exploration–exploitation policies that lack the capacity to adapt to shifting constraints in dynamic campaigns typical of self-driving laboratories. In this work, we develop a multi-objective BO framework for alloy design under resource constraints, benchmarking strategies for adaptive policy tuning at each iteration. Our evaluation covers a septenary refractory high-entropy alloy (RHEA) system focused on maximizing melting temperature and minimizing density, and a complex Fe-Co-Ni-based soft magnetic alloy system targeting the optimization of saturation magnetization, coercivity, and hardness. We compare an exploitation-focused strategy, a fixed mixed exploratory/exploitative policy, and two distinct LLM-based adaptive strategies with different approaches to batch allocation and campaign signal interpretation, evaluated across baseline and mid-campaign resource event conditions including budget reductions, timeline cuts, and combined disruptions. Our results show that mixed allocation strategies accumulate substantially more mutual information than the exploitation-focused baseline at a proportionally smaller cost to hypervolume and optimization speed, with adaptive strategies outperforming a fixed-mixed allocation policy by adjusting their allocation in response to both evolving campaign statistics and resource constraints. These findings suggest that adaptive resource allocation offers a favorable tradeoff for materials discovery campaigns in reducing predictive uncertainty on Pareto-optimal compositions.
    \end{abstract}


\end{frontmatter}

\section{Introduction}

Conventional experimentation in materials discovery and chemical sciences remains slow and resource-intensive, relying on sequential cycles of human intuition, hypothesis-driven testing, and trial-and-error optimization. As a result, the pace of discovery often falls short of the growing demand for advanced materials and sustainable chemical processes \cite{lee2026toward, bayley2024autonomous}. In response, self-driving labs (SDLs) and materials acceleration platforms (MAPs) have emerged as a new transformative paradigm of accelerated scientific discovery, integrating automation, artificial intelligence (AI), and algorithmic decision-making into closed-loop workflows to perform autonomous experiments with minimal human intervention \cite{lee2026toward, bayley2024autonomous, hickman2023self, hase2019next}. Within these systems, AI models are continuously updated using real-time experimental feedback to capture the behavior of the underlying system and guide the selection of subsequent experiments \cite{flores2020materials,roch2020chemos, sim2024chemos}. This closed-loop cycle, composed of experimentation, measurement, model refinement, and decision-making, enables data-efficient progress toward predefined objectives such as discovering new materials \cite{szymanski2023autonomous}, optimizing functional properties \cite{macleod2020self}, or improving reaction performance \cite{burger2020mobile}. In addition, the effectiveness of SDLs depends not only on individual components but also critically on the design, coordination, and adaptability of the overall experimental workflow \cite{hysmith2024future}.

Active learning has consequently become a central algorithmic component of SDLs, as closed-loop experimentation fundamentally requires selecting informative experiments under uncertainty. In a typical active-learning workflow, a surrogate model is iteratively updated, and its predictions and uncertainties are combined through an acquisition function to recommend experiments that balance exploration and exploitation \cite{lookman2019active}. In materials and chemical sciences, Bayesian optimization (BO), particularly multi-objective Bayesian optimization (MOBO), has emerged as a well-established realization of this framework, owing to the suitability of probabilistic surrogate models and exploration-exploitation strategies for settings characterized by expensive, noisy, and data-scarce experiments. 

Recent studies have demonstrated and benchmarked BO and MOBO across diverse discovery tasks, including multi-property materials design, establishing them as foundational tools for data-efficient experimentation \cite{liang2021benchmarking,wang2022benchmarking,arroyave2022perspective,adesiji2026benchmarking, desimpel2026bayesian, padhy2024experimentally}. In practice, these approaches are typically deployed in sequential or batched settings, where multiple experiments are proposed at each iteration to improve throughput, further increasing the complexity of decision-making.

Despite these advances, most implementations of BO in autonomous experimentation rely on fixed or manually tuned acquisition strategies \cite{de2021greed}. Such approaches implicitly assume stable experimental conditions and consistent objectives, whereas real-world experimental campaigns are often subject to changing constraints, including evolving design goals, limited budgets, and time-critical decisions. Under these conditions, the optimal balance between exploration and exploitation is inherently non-stationary, and fixed strategies can lead to inefficient allocation of experimental effort \cite{hoffman2011portfolio}. While BO provides statistically grounded mechanisms for selecting candidate experiments, it does not explicitly address the higher-level problem of adaptive policy selection, i.e., determining how the exploration–exploitation trade-off should evolve throughout the course of a campaign. This limitation can be viewed as a separation between local decision-making, governed by acquisition functions, and global decision-making, which governs how experimental effort is allocated over time \cite{armesto2023acquisition}. The absence of mechanisms to bridge this separation becomes particularly pronounced in batched and multi-objective settings, where trade-offs must be managed across competing objectives, parallel experimental allocations, and finite resource budgets~\cite{Daulton2020}.

In parallel with these developments, large language models (LLMs) have rapidly gained traction in chemical and materials research, primarily as interfaces for arduous tasks such as literature mining and knowledge extraction, planning, and tool-mediated assistance \cite{lei2024materials,jablonka202314,ramos2025review, jiang2025applications,dagdelen2024structured,katzer2025towards}. However, cautious assessments suggest that generic LLMs are not yet reliable domain reasoners and that effective deployment requires stronger grounding in domain-specific data, multimodal inputs, and structured interaction with expert tools \cite{miret2025enabling}. This perspective is supported by systems such as ChemCrow \cite{m2024augmenting}, which show that LLMs become substantially more useful when orchestrating search, safety, molecular, and reaction tools. CLAIRify \cite{yoshikawa2023large} and Coscientist \cite{boiko2023autonomous} demonstrate that language models can already participate in planning and execution layers of closed research workflows. Collectively, these studies suggest that LLMs are most effective as domain-aware coordinators that integrate prior knowledge, computational tools, laboratory systems, and human intent. This, in turn, raises the question of whether such systems can be extended to play structured roles in experimental decision-making.

In this context, a more concrete question is whether an LLM agent can dynamically manage a BO experiment, much like a human experimenter. Motivated by this perspective, recent work has begun to explore the role of LLMs in BO. For example, in chemistry, BoChemian \cite{rankovic2023bochemian} uses LLM embeddings of textual reaction procedures as feature representations for BO, whereas Kristiadi \etal \cite{kristiadi2024sober} takes a more cautious view in the discovery of molecular materials, showing that LLM-derived representations aid principled BO only when the underlying model is specific to the chemistry or adapted with domain data. In more general BO settings, LLAMBO \cite{liu2024large} leverages natural-language descriptions of prior evaluations for zero-shot warm starting, surrogate enhancement, and candidate sampling, highlighting the potential of language-based priors to influence early-stage search. 

More recent work has shifted from end-to-end proposal generation toward higher-level control. Gupta \emph{et al.} \cite{gupta2025llms} show that the open-source instruction-tuned LLM agents they evaluate are largely insensitive to experimental feedback and are consistently outperformed by linear bandits and Gaussian-process-based methods, motivating hybrid frameworks that separate prior-based reasoning from posterior-updating acquisition policies. In contrast, LMABO \cite{ngo2026adaptive} uses an LLM as a zero-shot strategist that selects among acquisition functions based on structured summaries of the optimization state, suggesting a more constrained and potentially more effective role for LLMs in guiding optimization. 

The findings described above point to a more structured role for LLMs—as bounded meta-decision layers that guide, rather than replace, established optimization frameworks. However, evidence for such roles in multi-objective optimization settings remains limited, particularly in scenarios where experimental constraints evolve dynamically and require adaptive reallocation of resources across competing objectives. This gap highlights the need for approaches that explicitly address decision-making at the level of policy selection, rather than candidate generation alone.

In this work, we address this gap by investigating whether an LLM-based agent can act as a high-level controller for batched MOBO. Specifically, we consider the problem of dynamically allocating experimental effort between exploration and exploitation under changing resource constraints, where the optimal strategy is inherently non-stationary. Our framework retains conventional MOBO components for surrogate modeling and batch acquisition, while introducing a separate decision layer that selects among structured allocation policies based on interpretable summaries of the optimization state, including campaign history, estimated hypervolume improvement (EHVI), information gain, and remaining budget and time. By explicitly separating policy selection from candidate generation, the proposed approach enables adaptive control over the exploration–exploitation trade-off while preserving the statistical efficiency and robustness of the underlying optimization framework. 

We evaluate this approach on two discrete alloy design problems formulated as multi-objective black-box optimization tasks over compositional spaces, and assess performance under simulated mid-campaign disruptions such as budget reductions and shortened timelines. By comparing against an exploitation-focused qEHVI baseline and a deterministic fixed mixed allocation strategy, we evaluate whether agent-guided allocation improves campaign responsiveness under evolving constraints. More broadly, this work establishes a generalizable framework for integrating probabilistic optimization with adaptive, language-driven decision-making, and highlights a pathway toward more flexible, scalable, and resilient self-driving laboratory systems operating under realistic, evolving constraints.

\section{Methods}
\label{sec:methods}

\begin{figure}
    \centering
    \includegraphics[width=1\linewidth]{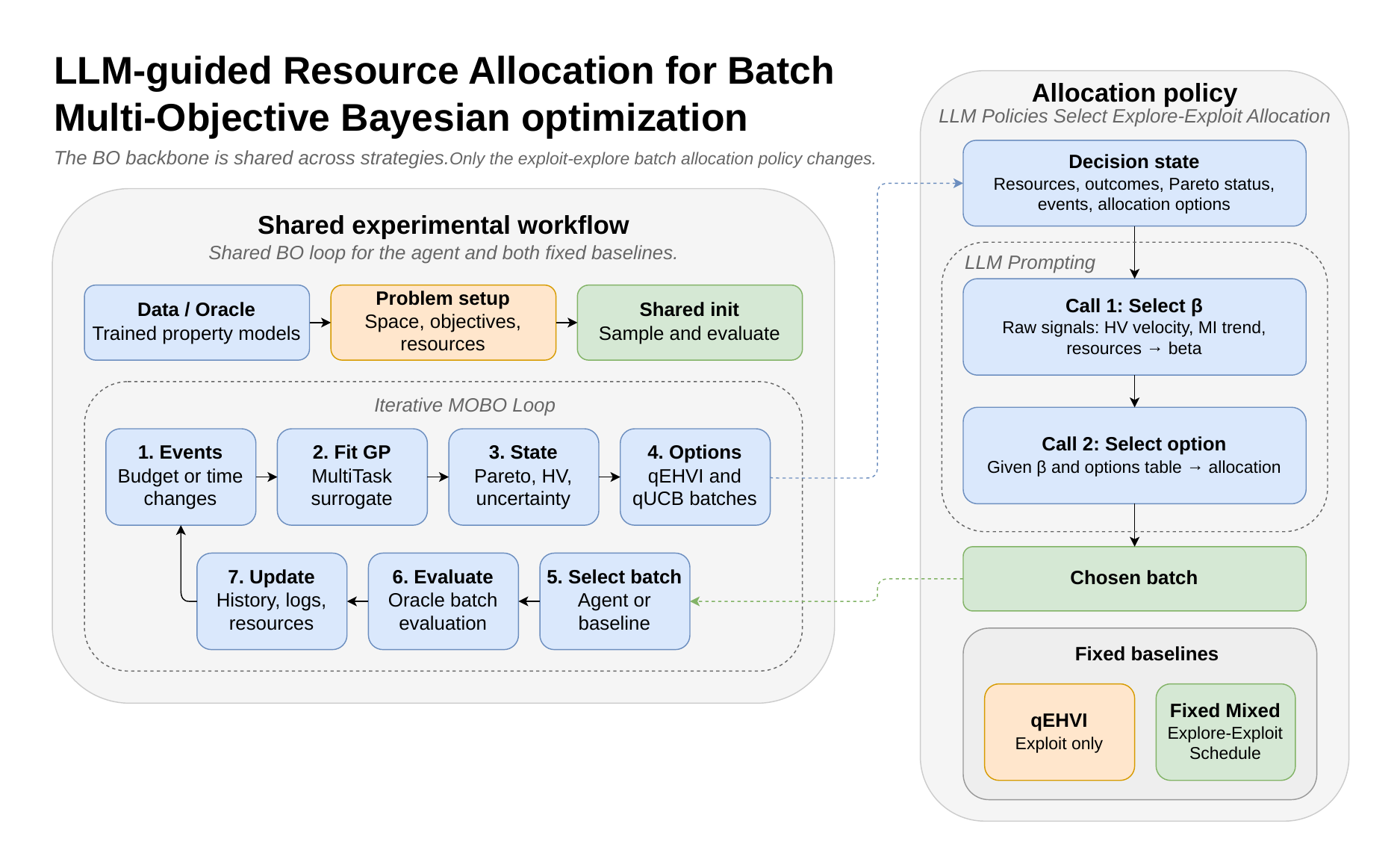}
    \caption{Overview of the developed LLM-guided resource allocation framework for batch multi-objective Bayesian optimization. The optimization loop is shared across all methods: surrogate fitting, state estimation, allocation-option generation, batch evaluation, and update. The proposed agent selects among exploit-explore batch allocations based on the current decision state, whereas the baselines use fixed qEHVI-only or qUCB-only allocations.}
    \label{fig:workflow}
\end{figure}

Here, we developed an agentic resource allocation framework to investigate dynamic exploration-exploitation balancing in batch multi-objective Bayesian optimization (MOBO) under evolving constraints. The framework comprises three integrated components: (i) high-fidelity ground truth models providing computationally tractable experiment proxies, (ii) a batch MOBO engine for candidate selection, and (iii) a resource allocation layer governing the exploration–exploitation balance at each iteration. We evaluate this framework on two multi-objective alloy composition optimization problems that serve as realistic test cases with competing objectives, discrete design spaces, and resource constraints representative of experimental materials discovery campaigns. Figure~\ref{fig:workflow} illustrates the overview of the agentic resource allocation framework and the interaction of these components within a complete optimization campaign.

\subsection{Test Problems: Alloy Composition Optimization}
\label{sec:problems}

The two systems evaluated comprised a three-objective problem in a high-dimensional multi-component soft-magnetic alloy composition space and a two-objective problem over a septenary refractory high-entropy alloy (RHEA) space.

The Fe-Co-Ni-based multi-component soft-magnetic alloy system was optimized by maximizing saturation magnetization $M_s$ in T, minimizing the logarithm of coercivity $\log_{10} H_c$ in A/m, and maximizing the logarithm of Vickers hardness $\log_{10} H_V$. High $M_s$ drives magnetic flux density, low $H_c$ reduces hysteresis losses, and sufficient $H_V$ ensures mechanical integrity, which together define the core trade-off surface for soft-magnetic applications such as transformers, inductors, and electromagnetic actuators~\cite{padhy2024experimentally}. The composition space comprises three primary elements (Fe, Co, Ni), each ranging over [0, 100] at.\%, subject to a group constraint requiring their sum to fall within [90, 100] at.\%, and six dopant elements (V, Mo, C, Nb, Ti, Si), each at most 5 at.\%, with their combined contribution constrained to at most 10 at.\%. All elements are sampled at 2 at.\% increments under these group constraints, yielding exactly 331,227 unique valid compositions. Training data were sourced from~\cite{padhy2024experimentally}. This literature-compiled database contains many compositions reported multiple times under different processing conditions, which composition-only inputs cannot distinguish, and its missing property values were imputed by Padhy \etal\ using property-specific machine learning models.

\label{sec:mpd-problem}
The septenary Ti-V-Nb-Mo-Hf-Ta-W refractory HEA (RHEA) system was optimized by maximizing melting temperature $T_m$ in K and minimizing density $\rho$ in g\,cm$^{-3}$ ~\cite{hastings2025leveraging}. High-temperature, low-density refractory alloys are of high interest for high-temperature structural applications such as aerospace propulsion applications~\cite{SENKOV20101758, ZHUO20241097}. Per-element bounds are Ti [0,35], V [0,50], Nb [22.5,67.5], Mo [0,12.5], Hf [0,12.5], Ta [0,45], W [0,12.5] at.\%, sampled at 5 at.\% spacing under the simplex constraint, yielding exactly 9,454 unique compositions.


\subsection{Ground Truth Models}
\label{sec:truth-models}

To enable the large-scale, multi-seed comparative study presented here, spanning hundreds of individual optimization campaigns, ground truth models were developed to serve as computational proxies for experimental evaluation. These ground truth models provide rapid and deterministic predictions for material properties. When the BO loop proposes a composition, the model returns objective values treated as exact experimental measurements.

Model construction was performed using ARMOTE-MO (Automated Regression workflow with Multi-Objective hyperparameter optimization using Tree-Parzen Estimator for Multi-Output regression), a framework that automates the end-to-end multi-output regression workflow, including data preprocessing, hyperparameter optimization, model training, and evaluation. ARMOTE-MO extends our previously developed automated regression workflow (ARW)~\cite{padhy2025automated,saurabh2026automated} to multi-output regression with multi-objective hyperparameter optimization. For both alloy systems, ARMOTE-MO trained a multi-output Random Forest Regressor (RFR)~\cite{Breiman2001} using scikit-learn~\cite{Pedregosa2011}, jointly predicting all objectives. A RFR model was selected owing to its strong baseline performance on tabular compositional data, computational tractability at the dataset scales used here ($10^4$--$10^5$ compositions), and native multi-output support without wrapper overhead. 

The framework manages train/test splitting (80/20) and standardizes inputs to zero mean and unit variance and scales outputs internally, ensuring fair comparison across properties with different physical units and magnitudes. Hyperparameters were optimized via multi-objective Bayesian optimization using Optuna~\cite{Akiba2019} with the Tree-Parzen Estimator (TPE), simultaneously minimizing MSE and maximizing $R^2$ across both outputs through 5-fold cross-validation on scaled split train dataset. This dual-objective search produces a Pareto front of hyperparameter configurations, from which the configuration with the highest mean $R^2$ across both objectives is selected.
The optimal hyperparameters for each system are listed in Table~\ref{tab:rfr-hyperparameters}. The complete ARMOTE-MO algorithm and full hyperparameter search spaces are provided in \ref{app:armote}.

\begin{table}[h]
\centering
\begin{tabular}{lcccc}
\toprule
System & $n_\text{trees}$ & max\_depth & min\_samples\_split & min\_samples\_leaf \\
\midrule
RHEA & 53 & 56 & 3 & 1 \\
Fe-Co-Ni        & 134 & 113 & 7 & 1 \\
\bottomrule
\end{tabular}
\caption{Optimal RFR hyperparameters for each alloy system selected by ARMOTE-MO.}
\label{tab:rfr-hyperparameters}
\end{table}

\begin{figure}[htbp]
    \centering
    \includegraphics[width=1\linewidth]{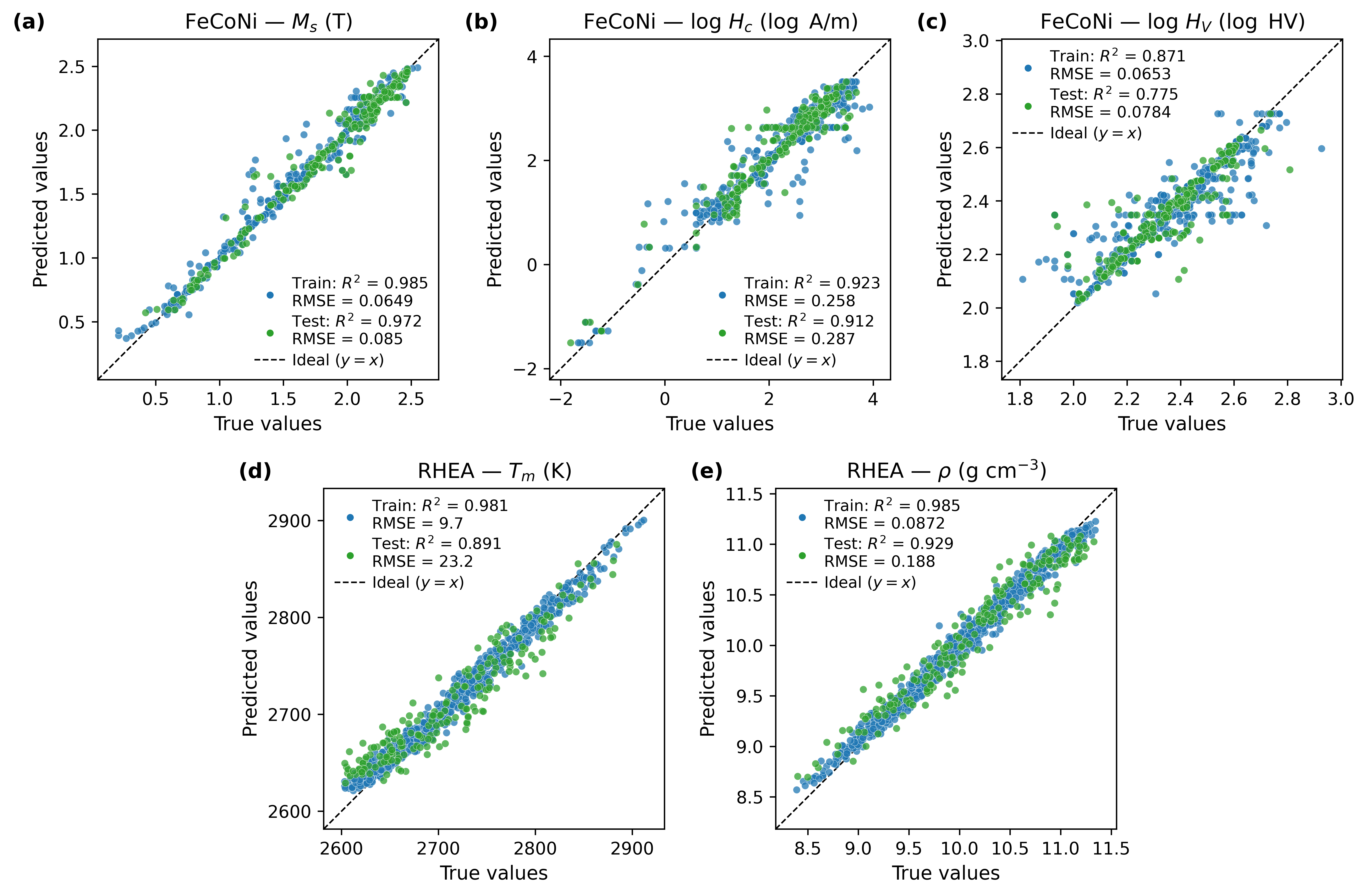}
    \caption{Parity plots for the ground truth random forest regression models. Panels (a)--(c) show the Fe--Co--Ni--V--Mo--C--Nb--Ti--Si soft magnetic system with predicted versus measured values for $M_s$ (T, $R^2 = 0.97$), $\log H_c$ ($\log(\text{A/m})$, $R^2 = 0.91$), and $\log H_V$ ($\log(\text{HV})$, $R^2 = 0.77$). Panels (d)--(e) show the Ti--V--Nb--Mo--Hf--Ta--W refractory system with predicted versus measured values for $T_m$ (K, $R^2 = 0.89$) and $\rho$ (g/cm$^3$, $R^2 = 0.93$). The dashed line indicates perfect agreement.}
    \label{fig:parity-plots}
\end{figure}

Both models achieve strong predictive accuracy, as confirmed by the parity plots in Fig.~\ref{fig:parity-plots}. Test $R^2$ values for the Fe-Co-Ni magnet system were $M_s$: 0.97, $\log H_c$: 0.91, and $\log H_V$: 0.77. For the RHEA system, $T_m$: 0.89 and $\rho$: 0.93. The lower $\log H_V$ accuracy reflects the greater scatter in hardness measurements across the dopant composition space, though the model captures the broad trends sufficient for comparative strategy evaluation. The slightly lower $T_m$ accuracy reflects the difficulty of predicting melting temperature across a wide compositional range, though the model captures the broad trends sufficient for comparative strategy evaluation. Prior to their use as inputs to the GP surrogate (Section~\ref{sec:gp}), all model outputs are standardized to improve numerical conditioning. Standardization statistics (per-objective mean and standard deviation) are estimated from 100 compositions sampled via quasi-random Sobol sequences from the enumerated design space and held fixed across all strategies and random seeds, ensuring that hypervolume values and GP length scales remain numerically consistent throughout the comparative study.

\subsection{Batch Multi-Objective Bayesian Optimization (MOBO) Engine}
\label{sec:bo-framework}

The batch MOBO engine is embedded within the broader resource allocation framework described in Section~\ref{sec:allocation-framework}. Each iteration begins with fitting a GP surrogate on accumulated data, constructing candidate batches with varying exploration–exploitation mixtures, selecting one batch via the resource allocator, evaluating the selected compositions, and updating the resource state.

All valid compositions are enumerated exhaustively at the specified grid spacings (Section~\ref{sec:problems}). Acquisition functions are optimized exactly over this finite discrete candidate set using BOTorch's~\cite{Balandat2020} \texttt{optimize\_acqf\_discrete}, eliminating continuous relaxation of the simplex constraint and ensuring grid-aligned proposals.

\label{sec:gp}
\texttt{MultiTaskGP}~\cite{Balandat2020,Bonilla2008,Swersky2013} from BOTorch is used to model both objectives jointly at each iteration, capturing inter-objective correlations through shared covariance. Observations from both objectives are interleaved into a single training set by appending a scalar task index to each input composition. Given $n$ evaluated compositions and $d$ elemental components, the training input matrix has shape $(2n,\, d+1)$ and the target vector has shape $(2n,\, 1)$, containing standardized objective values concatenated in task order. The GP hyperparameters are optimized by maximizing the marginal log-likelihood~\cite{Rasmussen2006} at each iteration. Input compositions are normalized to the unit hypercube based on per-dimension design space bounds prior to GP fitting and acquisition function evaluation.

\label{sec:acqf}
We use two acquisition functions to define the exploration–exploitation spectrum and serve as the building blocks for the allocation options:
\begin{itemize}

    \item Exploitation - qExpectedHypervolumeImprovement (qEHVI)~\cite{Daulton2020} measures expected increase in Pareto front hypervolume from a batch of $q$ compositions, estimated via a specified number (256 in our case) of Sobol quasi-Monte Carlo samples~\cite{Joe2008,Balandat2020}. It directs experiments toward known Pareto front improvements and thus operationalizes pure exploitation.

    \item Exploration - qUpperConfidenceBound (qUCB)~\cite{Wilson2017} balances high predicted values and high uncertainty via exploration coefficient $\beta=2.0$, applied to a linearly scalarized combination of both objectives with equal weights of 0.5. To prevent clustering with qEHVI candidates, we add a distance-based diversity bonus:
    \begin{equation}
    \label{eq:ucb-diversity}
    \text{UCB}_{\text{eff}}(\mathbf{x}) = \text{UCB}(\mathbf{x}) + \lambda \min_{\mathbf{x}' \in \mathcal{X}_{\text{qEHVI}}} \|\mathbf{x} - \mathbf{x}'\|_2
    \end{equation}
    where $\mathcal{X}_{\text{qEHVI}}$ are the set of candidates already selected by qEHVI in the current allocation option and $\lambda=1.0$ is the diversity weight. This drives qUCB toward regions of the design space that are geometrically distant from qEHVI candidates, ensuring that exploration and exploitation compositions cover complementary areas of composition space within each batch.

\end{itemize}

\label{sec:reference-point}
A fixed reference point $\mathbf{r}$ for hypervolume calculation is computed once by evaluating the full enumerated design space through the ground truth model and taking the per-objective minimum minus a margin of 0.1 in the post-processed objective space. This fixed $\mathbf{r}$ is shared across all strategies and seeds, ensuring that hypervolume values are directly comparable throughout the study.


\subsection{Resource Allocation Layer}
\label{sec:allocation-framework}

As mentioned in the Introduction, the core contribution is a resource allocation layer that determines how to split experimental effort between exploration and exploitation per batch at each iteration. Rather than fixing this split, the framework enumerates all feasible mixtures, characterizes each with exploration and exploitation metrics, and delegates selection to a resource allocator (fixed policy or LLM agent). This separates which compositions to evaluate from how to balance strategies over time, enabling dynamic policy control without modifying the MOBO engine.

Given a batch size $q$, we define $q+1$ allocation options indexed $i = 0, 1, \ldots, q$. Option $i$ allocates $q - i$ compositions to qEHVI (exploitation) and $i$ compositions to qUCB (exploration). Options are constructed sequentially. For option $i$, qEHVI first selects $q-i$ exploitation compositions; these are passed to qUCB as the anchor set $\mathcal{X}_\mathrm{qEHVI}$ in Eq.~\eqref{eq:ucb-diversity}, which then selects $i$ exploration compositions with diversity awareness. Therefore, option 0 is pure exploitation; option $q$ is pure exploration; intermediate options represent partial mixtures spanning the full exploration--exploitation spectrum. %
For both the RHEA and Fe-Co-Ni magnet experiments, we use $q = 5$, yielding six options (5+0, 4+1, 3+2, 2+3, 1+4, 0+5).

Each allocation option is characterized by two scalar metrics that together describe the expected utility from exploitation and exploration perspectives. These metrics are computed from the current GP surrogate before any evaluation occurs and are provided to the resource allocator as the basis for its decision.

\begin{itemize}

    \item Expected Hypervolume Improvement (EHVI) - The EHVI of an allocation quantifies the expected increase in the dominated hypervolume of the Pareto front if the combined batch were evaluated:
    \begin{equation}
    \label{eq:ehvi}
    \mathrm{EHVI}(X_S) = \mathbb{E}\!\left[
        \mathrm{HV}\!\left(\mathcal{Y}_\mathrm{obs} \cup f(X_S),\, \mathbf{r}\right)
        - \mathrm{HV}\!\left(\mathcal{Y}_\mathrm{obs},\, \mathbf{r}\right)
    \right]
    \end{equation}
    where $X_S$ is the $q$-composition batch, $f(\cdot)$ is the unknown objective function, $\mathcal{Y}_\mathrm{obs}$ is the set of all previously observed objective values, $\mathrm{HV}(\cdot, \mathbf{r})$ denotes the hypervolume dominated relative to reference point $\mathbf{r}$, and the expectation is taken over the GP posterior. EHVI is estimated by evaluating the qEHVI acquisition function on the combined batch $X_S$ using the current surrogate, regardless of which compositions in $X_S$ originated from qEHVI vs. qUCB. A high EHVI value indicates that the batch is well-positioned to substantially advance the known Pareto front.

    \item Mutual Information (MI) - The MI quantifies the expected reduction in uncertainty about the Pareto-optimal region of the objective space resulting from evaluating the batch $X_S$. Formally, it is the mutual information between the GP predictive distribution at $X_S$ and a sample from the surrogate Pareto-optimal set:

    \begin{equation}
    \label{eq:mi}
    \mathrm{MI}(X_S) = H\!\left(f(X_S)\right)
        - \frac{1}{\tilde{M}} \sum_{m=1}^{\tilde{M}} H\!\left(f(X_S) \mid f(\mathbf{x}^*_m)\right)
    \end{equation}
    where $\mathbf{x}^*_m$, $m = 1, \ldots, \tilde{M}$, are Monte Carlo samples from the surrogate Pareto-optimal set (we use $\tilde{M}$ here to distinguish the number of Pareto samples from the number of objectives $M$) and $H(\cdot)$ denotes differential entropy~\citep{Cover2006}. Under the Gaussian posterior of the MultiTaskGP, both entropy terms are computed analytically. For the joint distribution over $X_S$ evaluated at all $M$ objectives, the GP covariance matrix has dimension $Mq \times Mq$ ($M$ tasks, $q$ compositions per task). The prior entropy is:
    \begin{equation}
    \label{eq:prior-entropy}
    H\!\left(f(X_S)\right)
        = \frac{1}{2}\!\left[ Mq\ln(2\pi e) + \ln \left|\boldsymbol{\Sigma}_\mathrm{prior}\right| \right]
    \end{equation}
    where $\boldsymbol{\Sigma}_\mathrm{prior}$ is the $Mq \times Mq$ GP posterior covariance at the task-augmented inputs for $X_S$. After conditioning on a hypothetical observation at $\mathbf{x}^*_m$, the conditional entropy is:
    \begin{equation}
    \label{eq:cond-entropy}
    H\!\left(f(X_S) \mid f(\mathbf{x}^*_m)\right)
        = \frac{1}{2}\!\left[ Mq\ln(2\pi e)
            + \ln \left|\boldsymbol{\Sigma}_\mathrm{prior}
                - \boldsymbol{\Sigma}_{12}\boldsymbol{\Sigma}_{22}^{-1}\boldsymbol{\Sigma}_{21}
            \right| \right]
    \end{equation}
    where $\boldsymbol{\Sigma}_{12}$ is the $Mq \times M$ cross-covariance between $X_S$ and $\mathbf{x}^*_m$, $\boldsymbol{\Sigma}_{22}$ is the $M \times M$ posterior covariance at $\mathbf{x}^*_m$, and the Schur complement $\boldsymbol{\Sigma}_\mathrm{prior} - \boldsymbol{\Sigma}_{12}\boldsymbol{\Sigma}_{22}^{-1}\boldsymbol{\Sigma}_{21}$ is the conditional covariance of $f(X_S)$ given an observation at $\mathbf{x}^*_m$. A regularization term of $10^{-6}\mathbf{I}$ is added to all covariance matrices before computing log-determinants for numerical stability. The $\tilde{M} = 10$ surrogate Pareto samples are obtained by: (1) drawing 1000 compositions using Sobol sampling from the design space, (2) drawing a joint GP posterior sample path over the 1000 compositions for all $M$ objectives, (3) identifying the non-dominated set of these sampled objective vectors, and (4) drawing one composition uniformly from that set. This procedure is repeated $\tilde{M}$ times independently.

    MI is computed for each allocation option independently before any evaluation occurs, and we track both per-iteration and cumulative MI across the campaign as a measure of landscape characterization breadth. In real discovery campaigns, future constraints, objectives, or target regions are rarely fully known at the outset. A strategy that generates higher MI leaves behind a more thoroughly characterized landscape, broadening the range of Pareto-adjacent compositions that can be identified if campaign priorities shift. The default $\beta = 2$ used for qUCB tends to select compositions near the current Pareto boundary, where predicted values are high and uncertainty is moderate. When $\beta$ is elevated, proposals shift toward more speculative, high-uncertainty regions that are not necessarily close to the current front. MI remains a meaningful characterization measure in both regimes, as it quantifies the reduction in GP uncertainty about the Pareto-optimal region regardless of how close the proposed compositions are to the current front.

\end{itemize}

Within the resource allocation decision, EHVI values are compared across options to rank exploitation potential, and MI values are compared across options to rank exploration value. The resource allocator receives both metrics as inputs and is explicitly instructed to treat them as separate ordinal signals.

\subsection{Resource Allocation Strategies}
\label{sec:strategies}

We compared four resource allocation strategies that differ in how they select among the $q+1$ allocation options at each iteration. First are two fixed-policy baselines that require no campaign history.

The first strategy, called \textit{qEHVI} for simplicity, always selects option $i = 0$, directing the entire batch to the qEHVI acquisition function. This strategy serves as a reference for purely greedy acquisition behavior.

Next, the \textit{Fixed Mixed} strategy stages the allocation according to a deterministic schedule based on the fraction of resources remaining. When more than 75\% of total budget and time remain, it selects pure exploration (option $q$). For 25 to 75\% remaining, it selects the middle option, corresponding to 3 exploitation points and 2 exploration points for $q=5$. Below 25\%, it switches to pure exploitation (option 0). The exploration coefficient is fixed at $\beta = 2.0$ throughout. This policy captures a commonly applied experimental heuristic of exploring broadly early and then committing to the best-known region without requiring any learning or feedback.

The remaining two strategies are LLM-based agents designed to adapt to accumulated campaign evidence each iteration.

The \textit{Rule-Swap} strategy is an LLM-based agent whose primary objective is to maximize hypervolume as efficiently as possible, treating exploration as a targeted response to specific optimization failure modes rather than a routine investment at every iteration. The agent's primary diagnostic signal is hypervolume gain per newly added Pareto point (HV/pt) observed in recent qEHVI-dominant iterations. The agent defaults to pure qEHVI allocation and deviates only when the HV/pt trend shows crowding, where returns on new Pareto points are diminishing and qEHVI is refining known territory rather than expanding, or stagnation, where no new Pareto points have been added across several consecutive iterations. The degree of exploration added scales with the severity of the signal. The decision framing for when and how much to explore is encoded in the prompt rather than in Python control flow, so the LLM makes the actual judgment call within that framing. The agent makes two sequential LLM calls per iteration, with Call 1 producing a stagnation assessment and selecting $\beta$, and Call 2 selecting the allocation option using the resulting options table. Full prompt details are provided in \ref{app:prompts}.

\textit{Adaptive-$\beta$} takes a broader view of campaign state. Rather than a single diagnostic, it evaluates the HV velocity ratio (recent mean $\Delta$HV relative to the historical baseline), the current MI value relative to its value at the start of the campaign, per-element composition ranges across the current Pareto front to support domain reasoning about which regions of composition space have been covered, and a rolling six-iteration history of decisions and outcomes. Call 1 selects $\beta$ based on these signals. Call 2 then selects among the $q+1$ allocation options constructed at that $\beta$, guided by the Call 1 reasoning and the options table of qEHVI and MI scores. Unlike Rule-Swap, no fixed decision rules are encoded in the prompt. The LLM receives principled guidance text alongside raw computed signals and makes unconstrained decisions, with Python enforcement limited to output parsing and bounding the option index to $[0,q]$. All LLM calls for both agents use GPT-4o~\citep{OpenAI2024} (OpenAI API model identifier \texttt{gpt-4o}, queried 24--29 June 2026) through the LangChain~\citep{Chase2022} API. Full prompt details are provided in \ref{app:prompts}.

Resource events are mid-campaign modifications to budget or time constraints applied at a specified iteration, designed to simulate realistic laboratory disruptions such as unexpected funding reductions or shortened project timelines. The two fixed-policy baselines, qEHVI and Fixed Mixed, have no mechanism to respond and continue executing their schedules regardless of constraint changes. Both LLM-based agents receive notification of an event in the second LLM call at the iteration the event occurs, as well as a one-iteration advance warning in the preceding prompt (\ref{app:prompts_events}), giving them the opportunity to adjust allocation before the constraint takes effect.
 
\subsection{Experimental Design}
\label{sec:experimental-design}

Each optimization campaign begins with $n_\mathrm{init}$ compositions sampled uniformly at random from the enumerated discrete design space, evaluated through the ground truth model to produce the initial training dataset for the GP surrogate. The same $n_\mathrm{init}$ compositions and their corresponding objective values are shared across all four strategies within a given random seed, ensuring that differences in strategy performance reflect allocation decisions made during the campaign rather than differences in starting conditions. A new random initialization is drawn independently for each seed, with $n_\mathrm{init} = 15$ for the Fe-Co-Ni magnet experiments to accommodate the larger compositional search space and three-objective structure, and $n_\mathrm{init} = 5$ for the RHEA experiments.

All experiment configurations are summarized in Table~\ref{tab:experiments}. Each experiment is run with all four strategies using the same shared initialization per seed.

The two baseline experiments establish primary baselines comparing strategies in an unconstrained setting, while the six event experiments test strategy adaptability under mid-campaign resource variations where one or two constraint reductions are introduced during the early-to-mid phase of each campaign. These events are timed to fall when per-iteration gains are still substantial, creating the strongest test of whether the agent can recognize and respond to sudden changes in constraints. The baseline campaigns were designed such that the time and budget constraints will equally constrain the campaign. 

\begin{table}[h]
\centering
\begin{tabular}{llccccp{4.5cm}}
\toprule
Experiment         & System                     & Seeds & Iters & $q$ & $n_\mathrm{init}$ & Event(s) \\
\midrule
MP-D-MAIN     & RHEA & 25   & 55    & 5   & 5 & None \\
MP-D-BUDGET   & RHEA & 25   & 55    & 5   & 5 & Budget cut at iter.\ 15 \\
MP-D-TIME     & RHEA & 25   & 55    & 5   & 5 & Time cut at iter.\ 15 \\
MP-D-DUAL     & RHEA & 25   & 55    & 5   & 5 & Time cut at iter.\ 5; budget cut at iter.\ 7 \\
Magnets-MAIN   & Fe-Co-Ni & 25  & 50    & 5   & 15 & None \\
Magnets-BUDGET & Fe-Co-Ni & 25  & 50    & 5   & 15 & Budget cut at iter.\ 18 \\
Magnets-TIME   & Fe-Co-Ni & 25  & 50    & 5   & 15 & Time cut at iter.\ 13 \\
Magnets-DUAL   & Fe-Co-Ni & 25  & 50    & 5   & 15 & Time cut at iter.\ 13; budget cut at iter.\ 18 \\
\bottomrule
\end{tabular}
\caption{Summary of all experiment configurations. \textit{Iters} denotes the maximum number of BO iterations; campaigns with resource events terminate earlier when a constraint is exhausted. $q$ denotes batch size. $n_\mathrm{init}$ denotes the number of initial compositions shared across all strategies within each seed.
}
\label{tab:experiments}
\end{table}

The primary evaluation metric is the hypervolume indicator of the Pareto front approximation~\citep{Zitzler1998}. Given the fixed reference point $\mathbf{r} \in \mathbb{R}^M$ and the set of non-dominated evaluated compositions $\mathcal{Y}_\mathrm{pareto}$ accumulated through iteration $t$, the hypervolume is:
\begin{equation}
\label{eq:hypervolume}
\mathrm{HV}(\mathcal{Y}_\mathrm{pareto}, \mathbf{r})
    = \lambda\!\left(
        \left\{
            \mathbf{y} \in \mathbb{R}^M \;\middle|\;
            \mathbf{r} \preceq \mathbf{y}
            \text{ and }
            \exists\, \mathbf{p} \in \mathcal{Y}_\mathrm{pareto} : \mathbf{y} \preceq \mathbf{p}
        \right\}
    \right)
\end{equation}
where $\lambda$ denotes the Lebesgue measure ($M$-dimensional volume) and $\preceq$ denotes component-wise inequality. Larger hypervolume indicates the discovery of both high-performing and diverse alloys in the objective space. For the RHEA problem with $M=2$, the hypervolume is computed over melting point and negated density whereas in the Fe-Co-Ni magnet problem with $M=3$, it is computed over the three-dimensional objective space of $M_s$, $-\log H_c$, and $\log H_V$. In both cases, the fixed reference point described in Section~\ref{sec:reference-point} is used, enabling direct comparison across seeds and strategies within each system. 

To facilitate comparison across two systems with different hypervolume scales, we report hypervolume normalized by the theoretical maximum, specifically the hypervolume of the full design space Pareto front computed by evaluating the ground truth model over all enumerated compositions. This normalized HV lies in $[0, 1]$, where 1.0 corresponds to the best achievable Pareto front given the design space. Mean normalized HV and standard error of the mean across 25 random seeds are reported at each iteration for every strategy-configuration pair. As secondary metrics, the area under the HV curve (AUC-HV) is reported as a summary of early-campaign convergence speed, and cumulative mutual information in nats is reported as a measure of design space characterization breadth.

\section{Results}
\label{sec:results}

\subsection{Design Space and Objective Space Structure}
\label{sec:ground-truth-objectives}

\begin{figure}[htbp]
    \centering
    \includegraphics[width=0.9\linewidth]{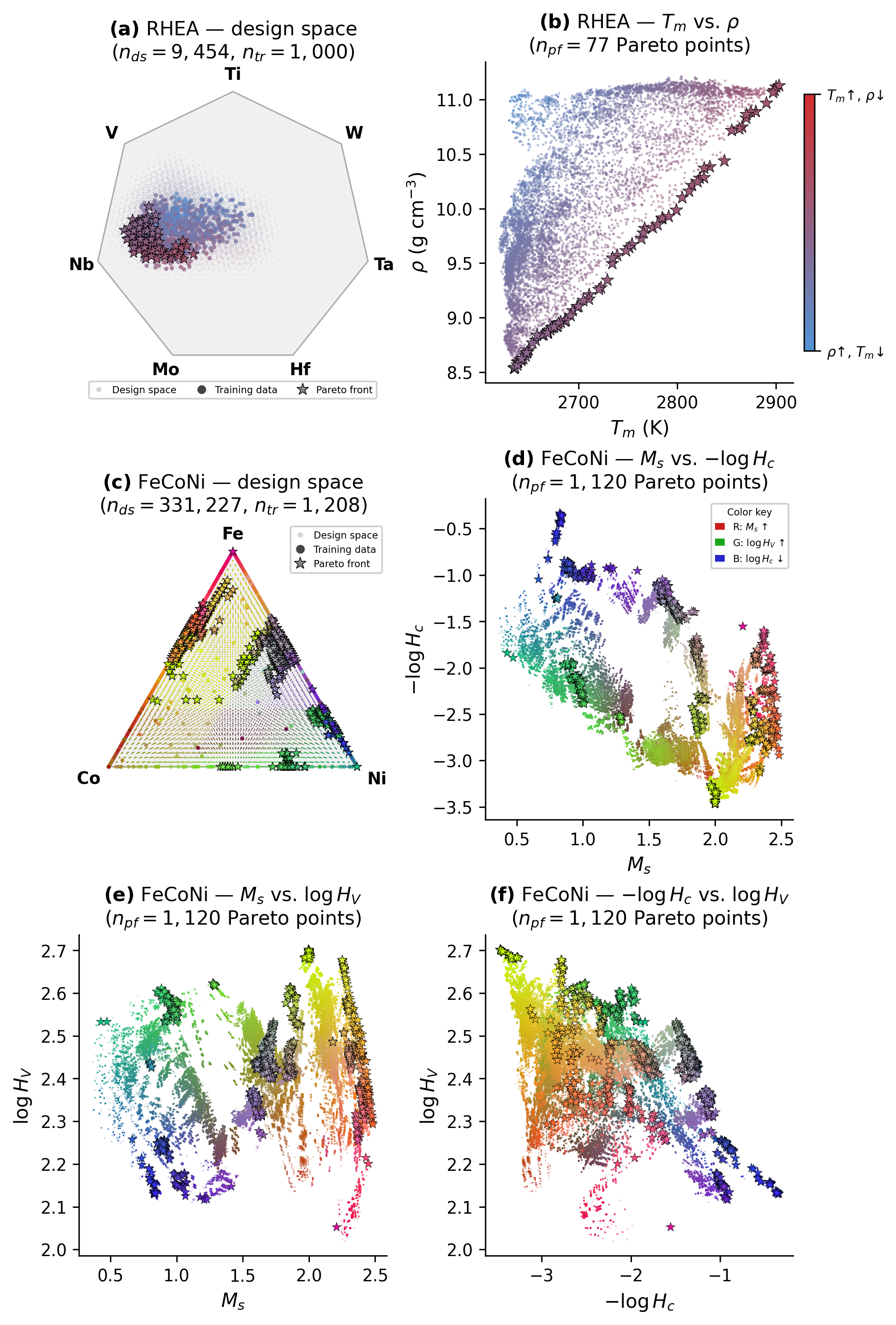}
    \caption{Ground-truth design and objective spaces for both systems.
             \textbf{(a)} RHEA system: training data and Pareto-optimal compositions shown in the septenary composition simplex.
             \textbf{(b)} RHEA objective space: density vs.\ melting temperature for all compositions, with Pareto-optimal points highlighted.
             \textbf{(c)} Fe-Co-Ni magnet system: training data and Pareto-optimal compositions projected onto the Fe--Co--Ni ternary, with Pareto-optimal compositions highlighted.
             \textbf{(d--f)} Fe-Co-Ni magnet objective space: pairwise projections of the three objectives, showing the multi-clustered structure of the Pareto front.}
    \label{fig:design-objectives}
\end{figure}

The RHEA and Fe-Co-Ni magnet problem design spaces were evaluated using the developed ground truth models to explore the actual Pareto fronts the strategies tried to discover. Fig.~\ref{fig:design-objectives} shows the composition and objective spaces for both systems.

The RHEA problem is focused on a relatively small contiguous cluster within the septenary simplex. Training data and Pareto-optimal compositions form a tight family of neighboring points in composition space, visible in panel (a). This spatial concentration carries over directly into the objective space in panel (b), where the density--melting temperature Pareto front spans a smooth and nearly linear tradeoff, with all optimal compositions grouped in a compact cluster. The optimization problem here is essentially one of locating and refining within a focused, well-connected region of the design space, a characteristic that distinguishes it from the Fe-Co-Ni magnet problem and will bear on interpreting the performance differences between them.

In contrast, the Fe-Co-Ni magnet problem looks quite different. Training data, shown in panel (c), is extensive along the binary edges of the ternary representation but becomes sparse in the interior. Despite this uneven coverage, the Pareto-optimal points clustered along specific binary edges and within isolated regions of the ternary interior, with clear discontinuities between groups. The three objective-space projections in panels (d)--(f) reinforce this picture, with every projection showing multiple distinct clusters of Pareto-optimal points rather than a continuous front. The optimization problem here involves finding and navigating a fragmented, multi-modal landscape substantially larger than the RHEA design space, representative of materials systems where the Pareto front is discontinuous and difficult to discover in full.

These two systems were chosen to represent qualitatively different optimization scenarios, with one presenting a compact and smooth landscape and the other a large and multi-clustered one, to stress-test fixed and adaptive resource-allocation strategies under different regimes. The experiments that follow probe whether strategy choice matters differently depending on which regime the campaign is operating in.
Using these ground truth models, we evaluate Adaptive-$\beta$, Rule-Swap, qEHVI, and Fixed Mixed across all eight experimental configurations described in Section~\ref{sec:experimental-design}, for both alloy systems.

\subsection{Baseline Optimization Performance}
\label{sec:results-baseline}

\begin{figure}[htbp]
    \centering
    \includegraphics[width=\textwidth]{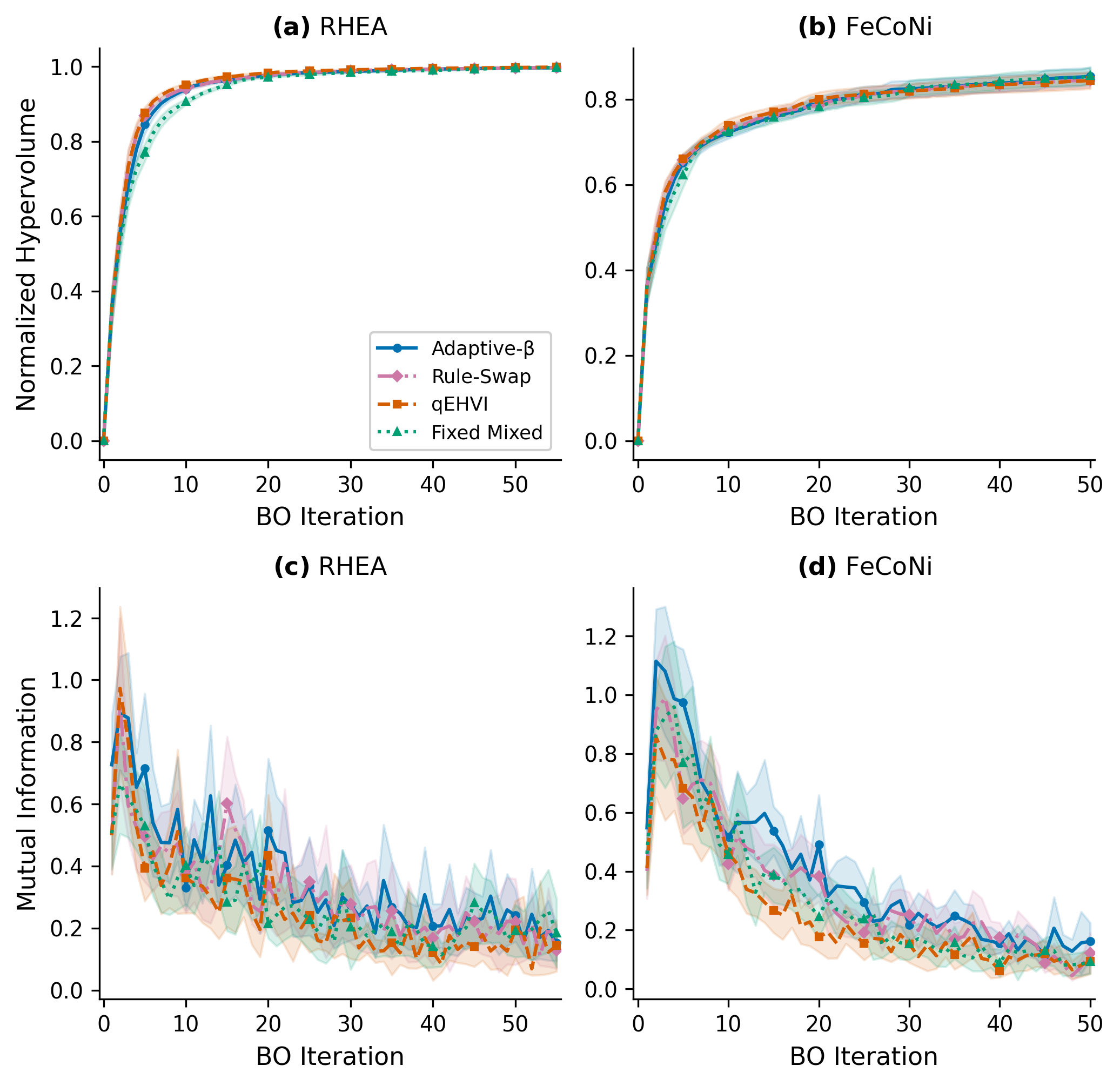}
    \caption{Baseline campaign results for both alloy systems with no resource events ($n = 25$ seeds per strategy).
             \textbf{a--b:} Hypervolume convergence; lines show per-strategy means and shaded bands indicate 95\% confidence intervals.
             \textbf{c--d:} Per-iteration mutual information across seeds; shaded bands are 95\% confidence intervals.
             Left column: RHEA system ($q = 5$, 55 iterations).
             Right column: Fe-Co-Ni magnet system ($q = 5$, 50 iterations).
             Strategies: Adaptive-$\beta$ (blue), Rule-Swap (purple), qEHVI (orange), Fixed Mixed (green).}
    \label{fig:baseline}
\end{figure}

\begin{figure}[htbp]
    \centering
    \includegraphics[width=\textwidth]{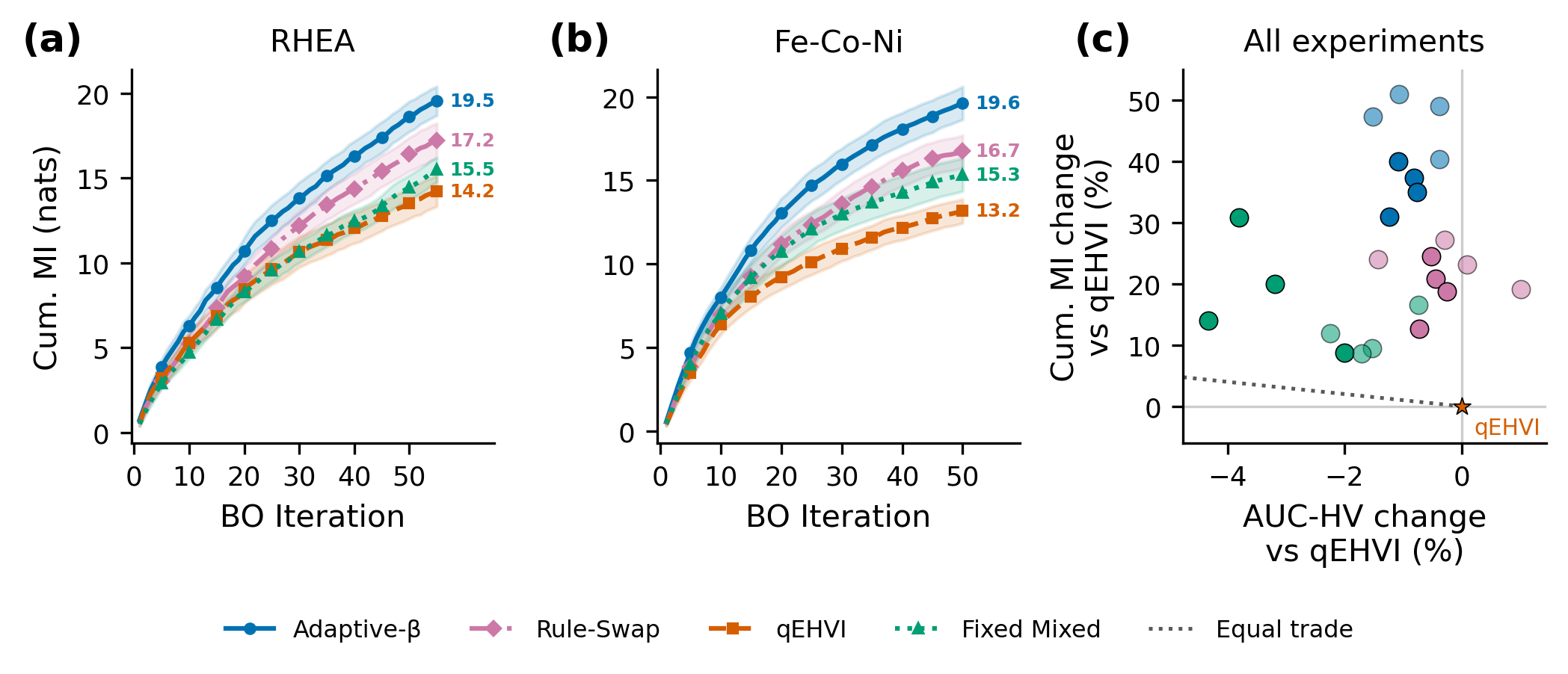}
    \caption{The information--optimization tradeoff across all eight experimental
             configurations ($n = 25$ seeds per strategy).
             \textbf{(a--b)} Cumulative mutual information over the baseline
             campaign for the RHEA and Fe-Co-Ni magnet systems; lines are
             per-strategy means and shaded bands are 95\% confidence intervals.
             \textbf{(c)} Cumulative MI gain plotted against AUC-HV cost, both
             expressed as a percentage change relative to the qEHVI baseline of
             the same experiment, which sits at the origin. Marker colour
             denotes strategy, and each experiment contributes one marker per
             mixed-allocation strategy; filled
             markers are the RHEA system and translucent markers the Fe-Co-Ni
             magnet system. The dotted line marks an equal trade, where the
             percentage of information gained equals the percentage of
             convergence speed given up. Every mixed-allocation point lies far
             above that line.
             End-of-campaign cumulative MI means are annotated at the
             right edge of (a) and (b); those values for all eight
             configurations, with their standard errors and alongside final
             hypervolume and AUC-HV, are tabulated in
             \ref{app:cumulative_mi_table}. Pairwise strategy
             comparisons reported anywhere in this work use the two-sided
             Wilcoxon signed-rank test on per-seed differences matched by
             shared random seed, with no multiple-comparison correction applied
             across the six strategy pairs within an experiment; the full
             matrix of $p$-values is given in
             \ref{app:significance}.}
    \label{fig:mi-tradeoff}
\end{figure}

In the RHEA system (Fig.~\ref{fig:baseline}, left column), all four strategies converged to nearly the same final hypervolume by the end of the 55-iteration baseline campaign. qEHVI reached the highest final hypervolume at $0.9983 \pm 0.00018$, followed by Fixed Mixed ($0.9971 \pm 0.00017$), Rule-Swap ($0.9970 \pm 0.00018$), and Adaptive-$\beta$ ($0.9967 \pm 0.00019$). Because the campaign was long enough for every strategy to approach the theoretical maximum, final hypervolume alone does not differentiate performance well. Examining how quickly each strategy accumulated the hypervolume over the course of the campaign reveals a clearer picture.
The area under the hypervolume curve (AUC-HV) showed a clear and consistent ordering in the RHEA system. qEHVI converged fastest at $52.02 \pm 0.08$, followed by Rule-Swap ($51.79 \pm 0.10$), Adaptive-$\beta$ ($51.59 \pm 0.09$), and Fixed Mixed ($50.98 \pm 0.09$). qEHVI, which allocates all resources toward exploitation of the predicted Pareto front, led throughout. The two adaptive strategies occupied the middle of the distribution, each accumulating roughly one fewer AUC unit than qEHVI over the 55-iteration campaign. Fixed Mixed fell furthest behind, reflecting its exploration-heavy early schedule that consistently delayed its ramp toward the high-performing region. This ordering follows how exploratory each strategy is early in the campaign, with greater early exploration yielding slower hypervolume accumulation.

The Fe-Co-Ni magnet system told a different story. The four strategies converged to mean final hypervolumes ranging from $0.84$ to $0.86$ over 50 iterations, with no statistically distinguishable differences between any strategy pair. AUC-HV values fell between $38.1$ and $38.4$ with no consistent ordering across strategies. The discontinuous Pareto front and larger design space likely drove the increased difficulty for all strategies. The near-identical outcomes across all strategies and seeds indicate that the Fe-Co-Ni magnet problem is simply very hard regardless of resource-allocation choice.

Mutual information separates the strategies where hypervolume does not. In both baseline campaigns, Adaptive-$\beta$ accumulated the most cumulative MI, followed by Rule-Swap, Fixed Mixed, and qEHVI; the per-configuration values for every experiment are given in Table~\ref{tab:cumulative-mi-full} and the baseline trajectories in Fig.~\ref{fig:mi-tradeoff}a--b. Adaptive-$\beta$ gained roughly $37\%$ more cumulative information than qEHVI over the 55 RHEA iterations and roughly $49\%$ more over the 50 Fe-Co-Ni magnet iterations. Its lead was established within the first ten iterations of both baseline campaigns and widened thereafter, so the separation is not an artifact of where a campaign happens to terminate.
\subsection{Optimization Performance Under Time and Budget Constraints}
\label{sec:results-events-hv}

\begin{figure}[htbp]
    \centering
    \includegraphics[width=\linewidth]{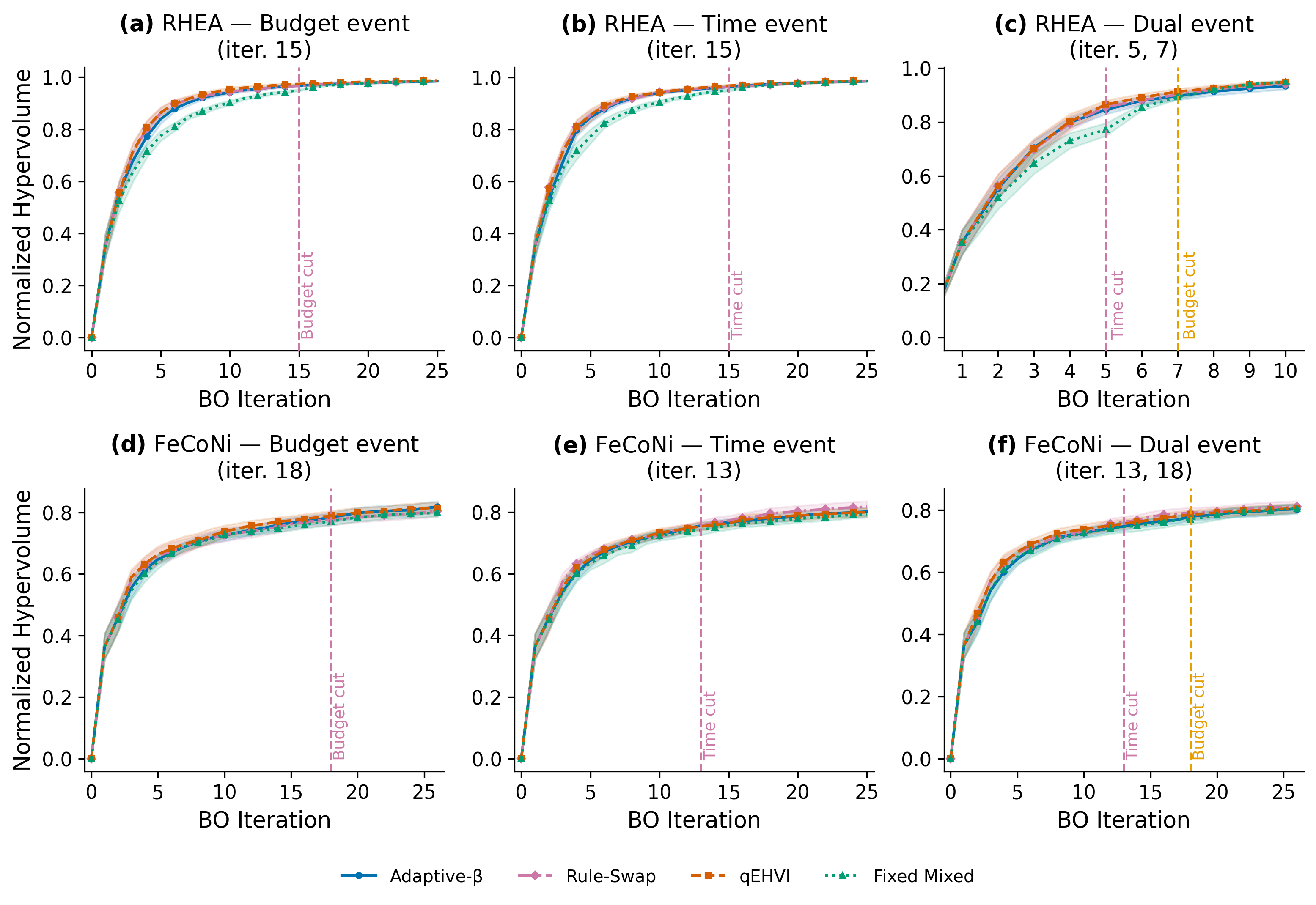}
    \caption{Hypervolume convergence for all six event experiments.
             Mean $\pm$ 95\% confidence interval across $n = 25$ seeds.
             Dashed vertical lines mark the iteration(s) at which resource
             events were applied.
             \textbf{Top row}: RHEA system ($q = 5$).
             \textbf{Bottom row}: Fe-Co-Ni magnet system ($q = 5$).
             From left to right: budget-event, time-event, and dual-event configurations.
             Strategies: Adaptive-$\beta$ (blue), Rule-Swap (purple), qEHVI (orange), Fixed Mixed (green).}
    \label{fig:event_hv}
\end{figure}

Fig.~\ref{fig:event_hv} shows hypervolume trajectories across all six event experiments. Before any event was applied, Adaptive-$\beta$, Rule-Swap, and qEHVI entered the event phase with comparable hypervolume accumulation across all six experiments. Fixed Mixed consistently entered the RHEA event experiments with lower pre-event AUC-HV than the other three strategies, the same consequence of its exploration-heavy early schedule seen at baseline. Its exploration period had not yet finished converting into hypervolume gains.

In the RHEA budget and time events, the separation between strategies widened compared to baseline. qEHVI again led in final hypervolume, reaching $0.9869 \pm 0.00063$ in the budget event and $0.9866 \pm 0.00100$ in the time event. Adaptive-$\beta$ finished significantly below qEHVI in both cases ($0.9843 \pm 0.00071$ budget, $0.9848 \pm 0.00054$ time), a gap that was not present at baseline where the two strategies were indistinguishable in final hypervolume. Rule-Swap remained comparable to qEHVI in both events ($0.9861 \pm 0.00048$ and $0.9858 \pm 0.00044$), and Fixed Mixed fell significantly below qEHVI in both ($0.9855 \pm 0.00043$ and $0.9846 \pm 0.00054$). For AUC-HV, the baseline ordering held. qEHVI led ($22.11 \pm 0.08$ budget, $22.00 \pm 0.09$ time), with Adaptive-$\beta$ and Rule-Swap following, and Fixed Mixed significantly below all other strategies in both events. Mutual information again separated the strategies, with Adaptive-$\beta$ accumulating the most in both events ($12.29 \pm 0.38$ budget, $11.97 \pm 0.37$ time) and all three mixed-allocation strategies significantly exceeding qEHVI ($8.78 \pm 0.31$ budget, $8.86 \pm 0.39$ time). Rule-Swap and Fixed Mixed were not distinguishable from each other in either event; the full comparison is given in Table~\ref{tab:cumulative-mi-full} and Table~\ref{tab:significance}.
The RHEA dual event represented an extreme case of campaign compression, truncating the total campaign to only ten iterations. Under these conditions, the strategy ordering for final hypervolume changed considerably from baseline. Fixed Mixed and qEHVI finished with comparable final hypervolumes ($0.9488 \pm 0.00216$ and $0.9486 \pm 0.00318$ respectively), together at the top of the distribution. Rule-Swap also finished comparably to both ($0.9413 \pm 0.00423$). Adaptive-$\beta$ fell significantly below both qEHVI and Fixed Mixed, finishing at $0.9341 \pm 0.00602$, the largest hypervolume spread across strategies in any experiment at approximately $1.5\%$ of the theoretical maximum compared to under $0.2\%$ at baseline. For AUC-HV, qEHVI led ($7.43 \pm 0.07$), with Adaptive-$\beta$ ($7.33 \pm 0.07$) and Rule-Swap ($7.37 \pm 0.07$) following and comparable to each other, and Fixed Mixed significantly below all three ($7.10 \pm 0.07$). The MI separation compressed over the ten remaining iterations but Adaptive-$\beta$'s lead persisted, at $6.11 \pm 0.35$ nats against $4.66 \pm 0.33$ for qEHVI. Rule-Swap ($5.25 \pm 0.26$) and Fixed Mixed ($5.32 \pm 0.18$) finished between the two and were indistinguishable from each other, so the middle of the baseline ordering no longer resolved. Adaptive-$\beta$ remained significantly above both Rule-Swap and qEHVI, and Rule-Swap remained significantly above qEHVI.

The Fe-Co-Ni magnet system produced largely indistinguishable results across all four strategies in final hypervolume and AUC-HV for all three event types, consistent with the baseline result. Strategies finished the budget event campaigns between $0.80$ and $0.82$ normalized hypervolume, the time event between $0.79$ and $0.82$, and the dual event between $0.80$ and $0.81$, with no consistent ordering in any case. One exception is the budget event, where qEHVI led Fixed Mixed in AUC-HV at $18.46 \pm 0.18$ versus $18.05 \pm 0.13$, and Adaptive-$\beta$ finished with a slightly higher final hypervolume than Fixed Mixed ($0.8186 \pm 0.0088$ versus $0.7992 \pm 0.0063$). In the time event, Rule-Swap led Fixed Mixed in AUC-HV at $17.62 \pm 0.12$ versus $17.17 \pm 0.16$. Mutual information across all three Fe-Co-Ni magnet event experiments followed the same ordering seen at baseline, with Adaptive-$\beta$ leading, Rule-Swap second, and both strategies accumulating substantially more MI than qEHVI.

\subsection{Convergence to Milestone Hypervolume Thresholds}
\label{sec:results-convergence}

\begin{table*}[]
\centering
\small
\setlength{\tabcolsep}{5pt}
\begin{tabular}{@{}lllccc@{}}
\toprule
\multirow{2}{*}{System} & \multirow{2}{*}{Experiment} & \multirow{2}{*}{Strategy}
  & \multicolumn{1}{c}{Batches to 50\% HV}
  & \multicolumn{2}{c}{Batches to 80\% HV} \\
\cmidrule(l){5-6}
 & & & Mean $\pm$ 95\% CI & Mean $\pm$ 95\% CI & \% reached \\
\midrule
  \multirow{16}{*}{Fe-Co-Ni} & \multirow{4}{*}{Baseline} & Adaptive-$\beta$ & $2.7 \pm 0.4$ & $29.7 \pm 4.6$ & 96\% \\
   &  & Rule-Swap & $2.5 \pm 0.4$ & $27.4 \pm 4.1$ & 92\% \\
   &  & qEHVI & $2.4 \pm 0.3$ & $26.3 \pm 4.2$ & 96\% \\
   &  & Fixed Mixed & $2.9 \pm 0.6$ & $28.3 \pm 3.9$ & 96\% \\
  \addlinespace[2pt]
   & \multirow{4}{*}{Budget event} & Adaptive-$\beta$ & $2.6 \pm 0.4$ & $20.1 \pm 2.5$ & 64\% \\
   &  & Rule-Swap & $2.6 \pm 0.5$ & $18.7 \pm 5.2$ & 40\% \\
   &  & qEHVI & $2.6 \pm 0.6$ & $18.6 \pm 4.4$ & 52\% \\
   &  & Fixed Mixed & $2.9 \pm 0.5$ & $20.0 \pm 3.6$ & 36\% \\
  \addlinespace[2pt]
   & \multirow{4}{*}{Time event} & Adaptive-$\beta$ & $2.8 \pm 0.5$ & $17.7 \pm 5.3$ & 28\% \\
   &  & Rule-Swap & $2.5 \pm 0.3$ & $17.5 \pm 3.3$ & 44\% \\
   &  & qEHVI & $2.6 \pm 0.4$ & $16.1 \pm 4.2$ & 32\% \\
   &  & Fixed Mixed & $3.0 \pm 0.5$ & $20.6 \pm 4.5$ & 32\% \\
  \addlinespace[2pt]
   & \multirow{4}{*}{Dual event} & Adaptive-$\beta$ & $2.9 \pm 0.5$ & $19.8 \pm 4.9$ & 36\% \\
   &  & Rule-Swap & $2.6 \pm 0.4$ & $16.6 \pm 3.7$ & 44\% \\
   &  & qEHVI & $2.7 \pm 0.4$ & $19.6 \pm 3.7$ & 52\% \\
   &  & Fixed Mixed & $2.9 \pm 0.5$ & $19.5 \pm 3.6$ & 52\% \\
\midrule
\multicolumn{4}{@{}l}{} & \multicolumn{2}{c}{Batches to 90\% HV} \\[-4pt]
\cmidrule(l){5-6}
  \multirow{16}{*}{RHEA} & \multirow{4}{*}{Baseline} & Adaptive-$\beta$ & $2.3 \pm 0.4$ & $7.4 \pm 0.6$ & 100\% \\
   &  & Rule-Swap & $2.2 \pm 0.3$ & $6.8 \pm 0.5$ & 100\% \\
   &  & qEHVI & $2.2 \pm 0.3$ & $6.2 \pm 0.5$ & 100\% \\
   &  & Fixed Mixed & $2.2 \pm 0.3$ & $9.8 \pm 0.6$ & 100\% \\
  \addlinespace[2pt]
   & \multirow{4}{*}{Budget event} & Adaptive-$\beta$ & $2.1 \pm 0.3$ & $7.3 \pm 0.5$ & 100\% \\
   &  & Rule-Swap & $2.1 \pm 0.3$ & $6.6 \pm 0.7$ & 100\% \\
   &  & qEHVI & $2.1 \pm 0.3$ & $6.6 \pm 0.7$ & 100\% \\
   &  & Fixed Mixed & $2.3 \pm 0.4$ & $10.1 \pm 0.5$ & 100\% \\
  \addlinespace[2pt]
   & \multirow{4}{*}{Time event} & Adaptive-$\beta$ & $2.4 \pm 0.4$ & $7.4 \pm 0.5$ & 100\% \\
   &  & Rule-Swap & $2.0 \pm 0.2$ & $7.0 \pm 0.6$ & 100\% \\
   &  & qEHVI & $2.0 \pm 0.2$ & $6.8 \pm 0.6$ & 100\% \\
   &  & Fixed Mixed & $2.2 \pm 0.3$ & $9.9 \pm 0.6$ & 100\% \\
  \addlinespace[2pt]
   & \multirow{4}{*}{Dual event} & Adaptive-$\beta$ & $2.2 \pm 0.3$ & $7.2 \pm 0.6$ & 88\% \\
   &  & Rule-Swap & $2.2 \pm 0.3$ & $7.2 \pm 0.5$ & 96\% \\
   &  & qEHVI & $2.2 \pm 0.3$ & $6.8 \pm 0.6$ & 100\% \\
   &  & Fixed Mixed & $2.4 \pm 0.4$ & $7.8 \pm 0.2$ & 100\% \\
\bottomrule
\end{tabular}
\caption{Batches required to reach 50\% and the upper milestone hypervolume threshold (mean $\pm$ 95\% CI; $n = 25$ seeds per strategy). All seeds reach the 50\% milestone in every experiment. Upper milestone thresholds differ by system, 80\% for the Fe-Co-Ni magnet problem; 90\% for the RHEA problem. The \% column reports the fraction of seeds reaching the upper milestone.}
\label{tab:convergence-speed}
\end{table*}

All four strategies reached the 50\% hypervolume milestone reliably in both systems and across every experiment. In the RHEA system, mean batch counts to the 50\% milestone were nearly identical across strategies and experiments, ranging from 2.0 to 2.4 batches, with all seeds reaching the threshold in every case. The 50\% threshold is reached early enough in RHEA campaigns that strategy choice has no meaningful effect on when it is first crossed. The Fe-Co-Ni magnet system required slightly more time at 2.4 to 3.0 mean batches at baseline, with Fixed Mixed taking about 0.5 more batches than qEHVI due to its exploration-heavy opening. Resource events shifted these early-campaign timings only negligibly in either system.

The 90\% milestone in the RHEA system revealed a clear and consistent ordering across the baseline, budget, and time event experiments. qEHVI reached the threshold fastest, averaging $6.2 \pm 0.5$, $6.6 \pm 0.7$, and $6.8 \pm 0.6$ batches in the baseline, budget, and time experiments respectively, all at 100\% seed coverage. Rule-Swap followed at $6.8 \pm 0.5$, $6.6 \pm 0.7$, and $7.0 \pm 0.6$ batches, and Adaptive-$\beta$ at $7.4 \pm 0.6$, $7.3 \pm 0.5$, and $7.4 \pm 0.5$ batches, both at 100\% coverage. Fixed Mixed was consistently the slowest at $9.8 \pm 0.6$, $10.1 \pm 0.5$, and $9.9 \pm 0.6$ batches, roughly 3--4 more than qEHVI in every case, reflecting the extended early exploration phase that delays its ramp toward the high-performing region. This ordering mirrors the AUC-HV pattern from Section~\ref{sec:results-baseline}, and budget and time events did not disrupt it.

The dual event was the one exception in the RHEA system. Fixed Mixed dropped from $9.8 \pm 0.6$ batches at baseline to $7.8 \pm 0.2$ batches at 100\% seed coverage, closing most of the gap to the other strategies. The dual event compressed the extended early exploration phase that normally slows Fixed Mixed down, forcing it into exploitation sooner. Adaptive-$\beta$ required $7.2 \pm 0.6$ batches at 88\% seed coverage, Rule-Swap $7.2 \pm 0.5$ batches at 96\% coverage, and qEHVI $6.8 \pm 0.6$ batches at 100\% coverage, all modest shifts within one batch of their baseline values. Notably, Adaptive-$\beta$'s coverage dropped to 88\%, meaning 3 of 25 seeds failed to reach the 90\% threshold within the compressed dual-event campaign.

The Fe-Co-Ni magnet system shows a different picture driven by problem difficulty. At baseline, all strategies reached the 80\% milestone in 92--96\% of seeds, with mean batch counts of 26.3 to 29.7 out of 50 total batches, meaning the upper milestone typically required more than half the entire campaign budget to reach. In budget events, coverage fell to 36--64\% in all strategies, with Adaptive-$\beta$ maintaining the highest coverage at 64\%, followed by qEHVI (52\%), Rule-Swap (40\%) and Fixed Mixed (36\%). Time events were more damaging, with coverage ranging from 28--44\% and no consistent ordering. Dual events produced similar coverage of 36--52\%, with qEHVI and Fixed Mixed both reaching 52\%. The combination of a large and fragmented design space with event-driven campaign truncation largely prevented strategies from reaching the upper hypervolume threshold in the Fe-Co-Ni magnet problem, making milestone comparisons between strategies unreliable under event conditions.

\subsection{\texorpdfstring{$\beta$}{β} Coefficient Trajectories Under Baseline and Event Conditions}
\label{sec:results-beta}

\begin{figure}[htbp]
    \centering
    \includegraphics[width=\textwidth]{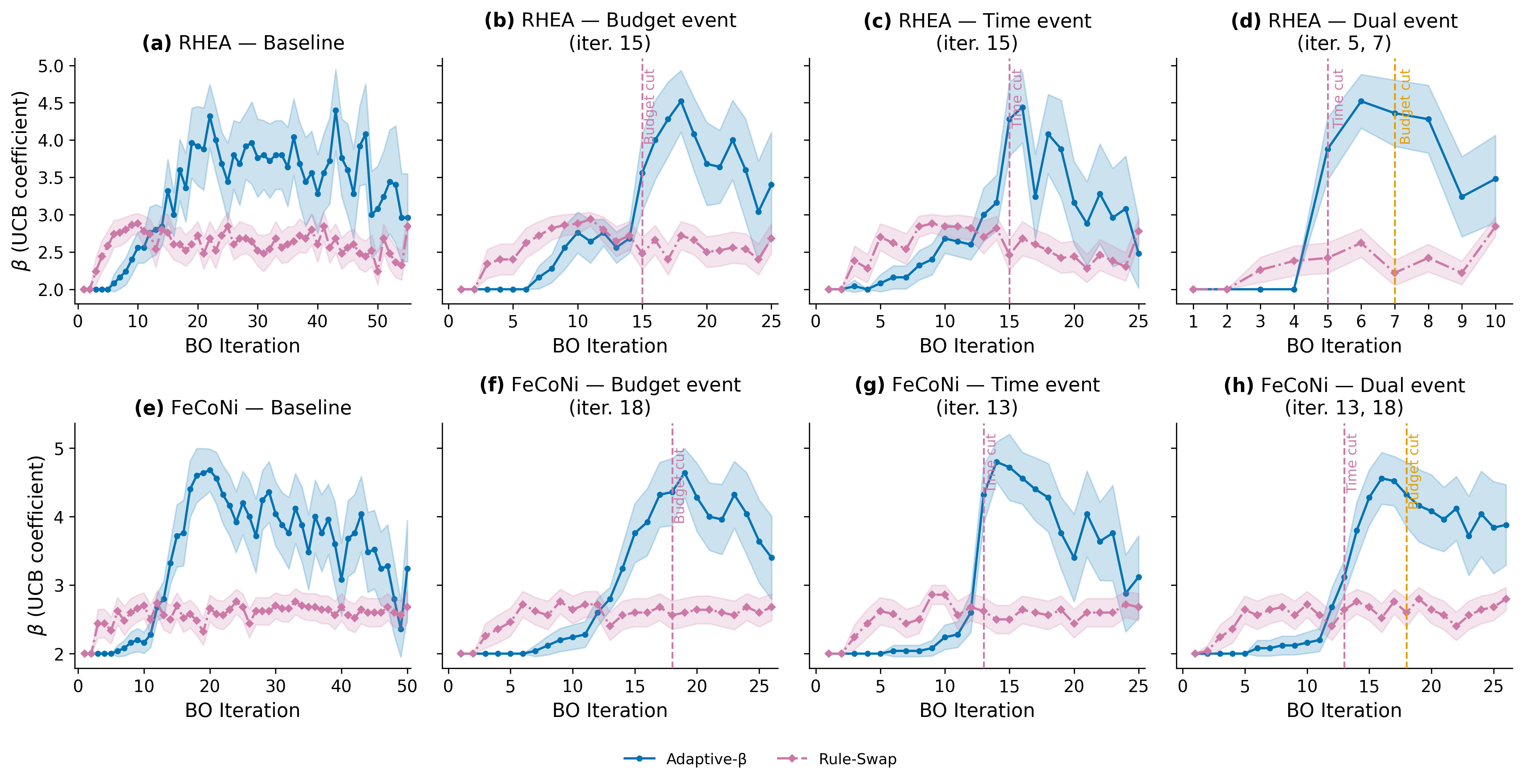}
    \caption{Adaptive-$\beta$ (blue, circles) and Rule-Swap (purple, diamonds) $\beta$ selection
             across all eight experimental configurations ($n = 25$ seeds per strategy).
             \textbf{Top row (a--d)}: RHEA system.
             \textbf{Bottom row (e--h)}: Fe-Co-Ni magnet system.
             Columns from left to right show the baseline, budget cut, time cut, and time-plus-budget cut configurations.
             Lines show per-seed means; shaded bands are 95\% confidence intervals.
             Dashed vertical lines mark the iterations at which resource events were applied.
             The dotted horizontal line marks $\beta = 2.0$, the minimum selectable value.}
    \label{fig:beta}
\end{figure}

The previous sections established how the strategies performed in terms of optimization outcomes. This section examines what the two LLM-based strategies were actually doing internally, specifically how Adaptive-$\beta$ and Rule-Swap set their parameters across the campaign and in response to events, in order to characterize whether their decision-making followed the intended design and what patterns emerged across problems.
At baseline, Rule-Swap's $\beta$ selections held within a narrow band centered near 2.5, spanning 2.0 to 3.5 without any directional trend in either system. This is consistent with the design of the rule shift, where $\beta$ is a secondary parameter, and the primary lever is the number of exploration candidates per batch. Adaptive-$\beta$ used $\beta$ as its main control mechanism and followed a very different trajectory. In the RHEA system, $\beta$ rose sharply from $2.14 \pm 0.03$ in early iterations (1--10) to $3.53 \pm 0.06$ in the middle phase (11--30), peaking near iteration 20, before plateauing through the final phase at $3.57 \pm 0.07$ (31--55). The Fe-Co-Ni magnet system followed a similar pattern, rising from $2.06 \pm 0.02$ in early iterations to $3.92 \pm 0.07$ in the middle phase before declining modestly to $3.55 \pm 0.08$ in the final phase. Both strategies became less exploratory as the campaign progressed, with Adaptive-$\beta$ doing so by ramping $\beta$ toward higher-variance candidate selection while gradually reducing exploration slots, and Rule-Swap by reducing its exploration candidate count toward zero in the final iterations of both baseline campaigns.

Looking at the within-campaign $\beta$ trajectory for Adaptive-$\beta$ in Fig.~\ref{fig:beta}, the ramp in the RHEA system is not smooth. The agent tended to make a large adjustment to $\beta$ and then hold for several iterations of smaller, stabilizing changes before making another large move, producing a step-like pattern. This is consistent with an agent that adjusts $\beta$ substantially, observes the effect on the noisy MI signal over a few iterations, and then re-adjusts. Given how much the per-iteration MI values fluctuate, as visible in Fig.~\ref{fig:baseline}, this test-and-adjust pattern is a plausible response to a feedback channel with high iteration-to-iteration variance.

Under event conditions, both strategies showed trends that were consistent with their baseline behavior, with the primary difference being timing. In the RHEA budget and time experiments, Adaptive-$\beta$'s post-event $\beta$ values were comparable to the baseline trajectory at the same iteration numbers, suggesting that the event advanced the campaign clock rather than triggering fundamentally new behavior. This was true regardless of whether the constraint was framed as a budget reduction or a time cut, indicating that the strategy response $\beta$ depended on the remaining iterations rather than the specific nature of the constraint. The RHEA dual experiment was the exception, where post-event $\beta$ significantly exceeded the baseline trajectory at those early iteration numbers, pointing to a genuine urgency response under extreme compression. In the Fe-Co-Ni magnet system, post-event $\beta$ matched the baseline trajectory across all three event types, mirroring the flat mid-to-late $\beta$ behavior seen at baseline. Rule-Swap's $\beta$ showed no consistent directional change under any event condition, and its primary behavioral response to events was in allocation, which is discussed in Section~\ref{sec:results-events-behavior}.

\subsection{Constraint Influence on Decision Making}
\label{sec:results-events-behavior}

\begin{figure}[htbp]
    \centering
    \includegraphics[width=\textwidth]{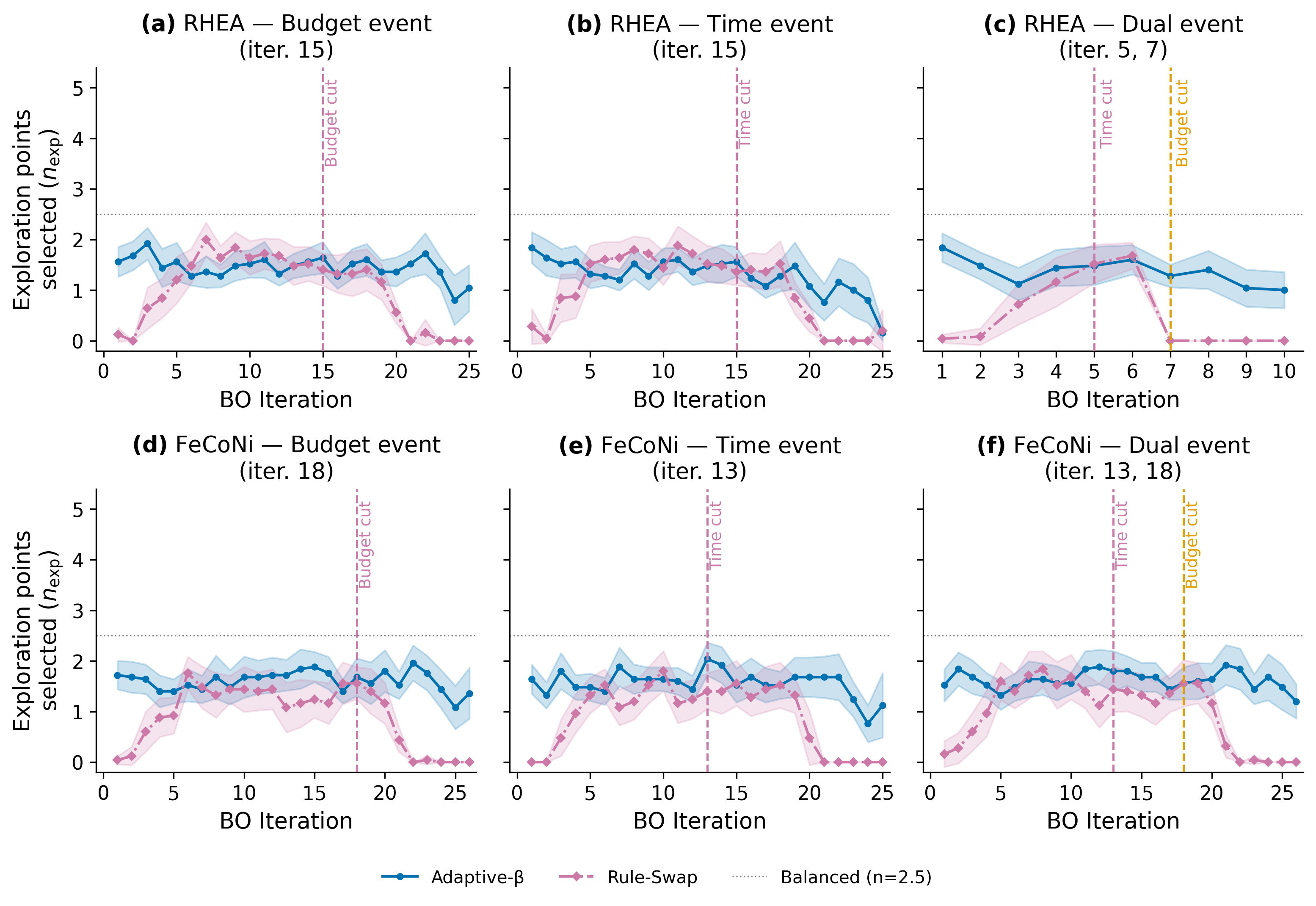}
    \caption{Exploration candidate allocation ($n_\mathrm{exp}$) for Adaptive-$\beta$ and Rule-Swap across all six
             event experiments.
             Mean across $n = 25$ seeds; shaded bands are 95\% confidence
             intervals. Dashed vertical lines mark the iteration at which
             each resource event was applied; colors distinguish event type
             (pink: first event; orange: second event in dual-event
             experiments).
             The dotted horizontal line marks the balanced allocation.
             \textbf{Top row}: RHEA system.
             \textbf{Bottom row}: Fe-Co-Ni magnet system.
             From left to right: budget-event, time-event, and dual-event configurations.}
    \label{fig:event_alloc}
\end{figure}

Fig.~\ref{fig:event_alloc} shows how Adaptive-$\beta$ and Rule-Swap adjusted their exploration candidate allocation ($n_\mathrm{exp}$) across all six event experiments. For Rule-Swap, $n_\mathrm{exp}$ was the primary mechanism through which the agent controlled resource allocation. Understanding how it changed with events therefore speaks directly to whether the strategy behaved as intended.
At baseline, both strategies selected approximately 1.5 exploration candidates per batch for most of the campaign in both systems. Rule-Swap gradually reduced $n_\mathrm{exp}$ toward zero in the final iterations of both baseline campaigns, consistent with its end-of-campaign shift toward exploitation. Adaptive-$\beta$ showed a similarly slight downward trend over time, generally maintaining at least one exploration candidate per batch throughout.

Under event conditions in the RHEA system, Rule-Swap's allocation response was immediate and large. In the budget event, post-event $n_\mathrm{exp}$ dropped to $0.59 \pm 0.06$ candidates per batch, compared to $1.37 \pm 0.04$ at the same baseline iterations, a reduction of $0.78$ candidates per batch. The time event produced a comparable reduction, with post-event $n_\mathrm{exp}$ of $0.58 \pm 0.05$ against a baseline of $1.37 \pm 0.04$, a drop of $0.79$ per batch. In the RHEA dual event, Rule-Swap's post-event $n_\mathrm{exp}$ reached zero and remained there for every remaining iteration.
Adaptive-$\beta$'s RHEA allocation response was substantially smaller and differed between event types. In the budget event, post-event $n_\mathrm{exp}$ was $1.36 \pm 0.06$, essentially unchanged from the $1.35 \pm 0.03$ at the baseline at the same iterations. The time event produced a larger reduction, with post-event $n_\mathrm{exp}$ dropping to $1.00 \pm 0.09$ from a baseline of $1.35 \pm 0.03$, a reduction of roughly $0.35$ candidates per batch. The dual event showed a modest but significant reduction from $1.36 \pm 0.03$ to $1.15 \pm 0.12$. Adaptive-$\beta$ treated a time constraint as a more direct reason to reduce exploration candidates than a budget constraint, suggesting that the framing of the constraint influenced the agent's allocation decisions differently even when both event types produced comparable $\beta$ responses.

In the Fe-Co-Ni magnet system, Rule-Swap again reduced $n_\mathrm{exp}$ significantly after each event type, dropping from $1.12 \pm 0.05$ to $0.38 \pm 0.05$ in the budget event, from $1.02 \pm 0.03$ to $0.75 \pm 0.06$ in the time event, and from $1.19 \pm 0.05$ to $0.39 \pm 0.04$ in the dual event, all relative to a baseline near $1.26$ at the same iterations. Adaptive-$\beta$ showed no significant change in $n_\mathrm{exp}$ after any of the three Fe-Co-Ni magnet event experiments, maintaining post-event allocations near $1.56$--$1.60$ candidates per batch, comparable to the baseline trajectory throughout. Rather than adjusting candidate counts in response to events, Adaptive-$\beta$'s primary adjustment mechanism in the Fe-Co-Ni magnet system was $\beta$, or in this case the absence of a $\beta$ response, as described in Section~\ref{sec:results-beta}. Across both systems and all event types, the two strategies responded to resource constraints through opposite mechanisms. Rule-Swap reduced exploration candidates consistently and aggressively, while Adaptive-$\beta$ held allocation steady and relied on $\beta$ as the primary adjustment lever.

\section{Discussion}
\label{sec:discussion}

In this work, with the set of experiments investigates, we show that the primary differences between resource allocation strategies lay in mutual information accumulated and convergence speed rather than final hypervolume, consistent across both alloy systems. No mixed-allocation strategy outperformed qEHVI for final hypervolume or AUC-HV, yet all three exceeded qEHVI substantially in cumulative MI. Fig.~\ref{fig:mi-tradeoff}c makes that tradeoff comparable across all eight configurations by expressing each mixed strategy as a percentage change against the qEHVI baseline of its own experiment, which removes the differing hypervolume and MI scales of the two systems. All twenty-four mixed-allocation points lie above the equal-trade line, where the percentage of information gained would equal the percentage of convergence speed given up, and they lie well above it: the median ratio of proportional information gained to proportional AUC-HV given up is roughly $28$, with no configuration below $3$. Adaptive-$\beta$ sits furthest into that region in both systems, and the Fe-Co-Ni magnet configurations cluster near the vertical axis, where the AUC-HV cost is small and, in all but one comparison, not statistically resolvable. The annotated baseline curves in Fig.~\ref{fig:mi-tradeoff}a--b and the full values in Table~\ref{tab:cumulative-mi-full} show the same qualitative ordering: Adaptive-$\beta$ accumulated the most cumulative MI, qEHVI the least, and Rule-Swap and Fixed Mixed fell between them. All four strategies produced statistically indistinguishable final hypervolumes on the Fe-Co-Ni magnet problem, making cumulative MI and AUC-HV the only metrics by which strategies could be meaningfully compared there.

The mixed-allocation strategies traded AUC-HV and convergence speed for greater uncertainty reduction around the Pareto front at each iteration. The greater MI gains suggested that the GP surrogate was likely better characterized in the Pareto-front neighborhood for exploration-biased campaigns than for purely exploitative ones, as higher MI per iteration reflected the acquisition of more informative measurements relative to the current surrogate uncertainty about the Pareto front. Whether that improved characterization translated into practically meaningful outcomes for downstream materials analysis, such as more informative experimental records for objective refinement or follow-on campaigns, was not directly measured here and remains an open question.

Whether this tradeoff favors an adaptive policy over a fixed mixed schedule is what the resource-constraint experiments reveal. Resource events did not substantially change per-iteration optimization performance for any strategy relative to the baseline at the same point in the campaign, as reflected in the milestone convergence data of Table~\ref{tab:convergence-speed}, where batch counts to the 90\% hypervolume threshold fell within one batch of baseline values for all budget and time events. The primary effect of events was to reveal differences in how strategies compared to one another over a shortened campaign. HV performance in the Fe-Co-Ni magnet problem was largely indistinguishable across strategies under single events. The budget event yielded two exceptions: Adaptive-$\beta$ finished above Fixed Mixed in final hypervolume ($0.8186 \pm 0.0088$ versus $0.7992 \pm 0.0063$, $p = 0.045$), and qEHVI led Fixed Mixed in AUC-HV ($18.46 \pm 0.18$ versus $18.05 \pm 0.13$, $p = 0.026$). In the time event, Rule-Swap also led Fixed Mixed in AUC-HV ($17.62 \pm 0.12$ versus $17.17 \pm 0.16$, $p = 0.027$). All other pairings remained statistically indistinguishable. 

However, the RHEA experiments showed a clearer effect of the shortened campaign. qEHVI required no behavioral response because it already allocated all resources to exploitation, and Rule-Swap compressed its end-of-campaign exploration allocation in response, while Adaptive-$\beta$'s allocation remained near baseline under the budget constraint and reduced more substantially under the time constraint. Fixed Mixed however could not adapt the mixed allocation according to the shortened campaign, causing the AUC-HV gap relative to the other three strategies to widen across all three RHEA event types compared to baseline, visible in Fig.~\ref{fig:event_hv}. This was a consequence of its committed early exploration schedule that couldn't be converted into hypervolume gains by the time the event arrived. Adaptive-$\beta$ retained its cumulative-MI lead over qEHVI under every event condition in both systems. The RHEA dual event produced the most pronounced shift in relative performance visible in Fig.~\ref{fig:event_hv}. 

Two consecutive resource events reduced the RHEA campaign to ten total iterations, and Adaptive-$\beta$ fell significantly below both qEHVI and Fixed Mixed in final hypervolume, finishing at $0.9341 \pm 0.00602$ compared to $0.9488 \pm 0.00216$ for Fixed Mixed and $0.9486 \pm 0.00318$ for qEHVI. This was a reversal from the baseline, where Adaptive-$\beta$ and Fixed Mixed were statistically indistinguishable in final hypervolume ($p = 0.080$). The dual event simultaneously narrowed the milestone gap for Fixed Mixed in Table~\ref{tab:convergence-speed}, which reached the 90\% threshold in $7.8 \pm 0.2$ batches compared to $9.8 \pm 0.6$ at baseline, because it compressed the extended early exploration phase that normally delayed it. Adaptive-$\beta$ dropped to 88\% milestone coverage with three seeds failing to reach the threshold, because ten iterations was insufficient for Adaptive-$\beta$ to accumulate the campaign history that evidence-based decisions required. The absolute differences remained below 2\% of the theoretical maximum, but the dual event suggested that below some campaign-length threshold, evidence-based allocation strategies could begin to fall behind strategies that do not require history to act.

Both LLM-based strategies adopted broadly consistent behavioral patterns throughout the campaign and adjusted the timing of those patterns rather than the patterns themselves when resource events occurred, as seen across all panels of Fig.~\ref{fig:beta} and Fig.~\ref{fig:event_alloc}. Rule-Swap followed hard-set rules designed to introduce exploration when qEHVI performance was slowing. It maintained beta selections near 2.5 throughout both baseline campaigns, and gradually reduced $n_\mathrm{exp}$ toward zero at the end of each campaign. Events moved the point at which that end-of-campaign reduction occurred without changing its form. 

Adaptive-$\beta$ selected approximately 1.5 exploration candidates per batch on average, slowly decreasing across the campaign, while beta followed a more structured trajectory. Beta ramped in the early phase of both campaigns, reaching a peak near iteration 20 (approximately 36\% of the way through the RHEA campaign and 40\% through the Fe-Co-Ni magnet campaign), after which it followed a cyclical pattern of larger adjustments followed by smaller stabilizing changes with the overall envelope slowly declining toward the end of the campaign. The cyclical pattern is likely due to the fixed number of iterations that are permitted to exist in the agent's short term memory. The overall shift away from exploration as the campaign progressed was likely driven by the diminishing MI gain available at each iteration as the surrogate matured, combined with fewer remaining iterations to recover from any exploratory deviation. Both agents responded to events by advancing their trajectory rather than departing from it, suggesting both had converged on an allocation approach that adapted its pace rather than its overall strategy.

The type of constraint applied during events, whether a budget cut or a shortened timeline, did not produce different beta trajectories for either strategy, as seen in Fig.~\ref{fig:beta}. Rule-Swap's allocation showed no meaningful difference between constraint types either, reducing $n_\mathrm{exp}$ from $1.37 \pm 0.04$ at comparable baseline iterations to $0.59 \pm 0.06$ under a budget constraint and $0.58 \pm 0.05$ under a time constraint. Adaptive-$\beta$'s allocation was slightly lower under time constraints than budget constraints ($1.00 \pm 0.09$ versus $1.36 \pm 0.06$ candidates per batch, visible in Fig.~\ref{fig:event_alloc}), suggesting that framing the constraint as a deadline had a small but distinguishable effect on its candidate counts relative to an equivalent budget reduction.

The behavioral patterns described above emerge from a deliberate design tradeoff. Restricting both agents to optimization statistics only (hypervolume history, cumulative MI, and remaining campaign resources) limited the range of behaviors they could explore, but those constraints were required to enforce evidence-based decisions without introducing problem-specific bias. Augmenting the agents' decision context with prior information regarding the material family being optimized could improve early-campaign allocation before sufficient optimization history had accumulated.

\section{Conclusion}

We investigated how mixed and adaptive resource allocation strategies influence the exploration--exploitation tradeoff in batch multi-objective Bayesian optimization under evolving resource constraints in autonomous materials discovery campaigns. Fixed, rule-based adaptive, and LLM-guided adaptive policies were evaluated across a three-objective soft-magnetic alloy design problem ($M_s/H_c/H_V$ in the Fe--Co--Ni--V--Mo--C--Nb--Ti--Si system) and a two-objective melting-point/density problem ($T_m/\rho$ in the Ti--V--Nb--Mo--Hf--Ta--W system), spanning eight experimental configurations covering baseline and single- and dual-event conditions.
The central quantitative result is that the percentage gain in cumulative mutual information achieved by mixed allocation strategies substantially exceeded the corresponding percentage loss in hypervolume and AUC-HV relative to qEHVI. 

In the RHEA baseline campaign, Adaptive-$\beta$ accumulated approximately 37\% more cumulative MI than qEHVI ($19.54 \pm 0.41$ versus $14.23 \pm 0.45$ nats) while finishing within 0.2\% of qEHVI in final hypervolume and within 0.8\% in AUC-HV. In the Fe-Co-Ni magnet problem, where all strategies produced statistically indistinguishable final hypervolumes and AUC-HV values, Adaptive-$\beta$ accumulated approximately 49\% more MI than qEHVI at no measurable cost in optimization performance. Rule-Swap and Fixed Mixed fell between these values in the same ordering across both systems. 

The implication is that exploration-biased campaigns acquired a substantially more informative characterization of the Pareto-front neighborhood even when optimization convergence appeared uniform, a distinction whose practical value lies in the downstream utility of the experimental record for objective refinement and follow-up work.

Among the mixed allocation strategies, the adaptive policies outperformed Fixed Mixed in responsiveness to both evolving campaign statistics and mid-campaign resource events, advancing their normal allocation trajectory in response to constraints rather than altering the underlying strategy itself, while Fixed Mixed's predetermined exploration schedule could not be revised to account for the shortened campaigns.

The primary limitation of the current framework is that both LLM-based strategies operated exclusively on optimization statistics, drawing on hypervolume history, cumulative MI, and remaining campaign resources without access to any prior knowledge about the materials system being studied. This restriction enforced evidence-based decisions and avoided problem-specific bias, but it also meant that allocation decisions in the early campaign, before sufficient optimization history had accumulated, were made without knowledge of how composition relates to the objectives being optimized.

These findings point toward several directions for improving LLM-guided allocation agents in self-driving laboratory settings. Resource allocation agents equipped with prior information about a materials system could influence the selection of candidate points beyond simply choosing among acquisition function suggestions, directing early-campaign sampling toward regions of composition space that prior knowledge identifies as more likely to yield Pareto-relevant compositions. 

A further open question is the extent to which the mutual information advantages documented across all experiments translate into measurable improvements in Pareto-front prediction accuracy. Quantifying this relationship would provide a principled basis for selecting the exploration intensity that maximizes the long-term utility of an experimental campaign rather than optimizing solely for hypervolume within a fixed horizon. More broadly, this work establishes a generalizable framework for integrating probabilistic optimization with adaptive, language-driven decision-making, and demonstrates that LLM agents can play a structured and beneficial role as high-level controllers in multi-objective experimental campaigns operating under realistic, evolving constraints.
\section*{Conflicts of interest}
There are no conflicts to declare.

\section*{Data availability}
Data for this article, including the ground truth models and their training data, the enumerated design spaces, and the complete results of all optimization campaigns (per-iteration convergence, evaluations, candidate allocation options, and LLM agent logs), are available on GitHub at \url{https://github.com/Tripp5118/resource-allocation}. The code for the resource allocation framework, the ground truth model workflow, and the analysis scripts that reproduce all figures and tables in this article can be found at the same location. The version of the code and data employed for this study is version v1.0.1, archived on Zenodo~\cite{robinson2026agentic} at \url{https://doi.org/10.5281/zenodo.23111493}. The Fe-Co-Ni training data were obtained from Padhy \etal~\cite{padhy2024experimentally, padhy2024data}, available at \url{https://doi.org/10.5281/zenodo.10686272}, and the Ti-V-Nb-Mo-Hf-Ta-W training data from Hastings \etal~\cite{hastings2025leveraging}.

\section*{Acknowledgements}
The research was sponsored by the Army Research Laboratory and was accomplished under Cooperative Agreement Number W911NF-22-2-0106. The views and conclusions contained in this document are those of the authors. They should not be interpreted as representing the official policies, either expressed or implied, of the Army Research Laboratory or the US Government. The US Government is authorized to reproduce and distribute reprints for Government Purposes, notwithstanding any copyright notation herein. Original data were generated within the BIRDSHOT Center (https://birdshot.tamu.edu), supported by the Army Research Laboratory under Cooperative Agreement (CA) Number W911NF-22-2-0106 (All Authors acknowledge support from this CA). RA also acknowledges NSF through Grant No. DMR-2534344 (Autonomous Robotic Metallurgist Materials Innovation Platform, ARM-MIP).

We also acknowledge the use of OpenAI's ChatGPT-5 and Anthropic's Claude Sonnet-4 for some sentence structure and code structure formations in the manuscript, respectively. After using this tool/service, the author(s) reviewed and edited the content as needed and take(s) full responsibility for the code and content of the publication. 

\nolinenumbers

\bibliography{mybibfile,extra}

\clearpage
\appendix
\definecolor{promptblue}{HTML}{EAF2FF}
\definecolor{promptgreen}{HTML}{ECF8F1}
\definecolor{promptorange}{HTML}{FFF4E8}
\definecolor{frameblue}{HTML}{2E5AAC}
\definecolor{framegreen}{HTML}{2F7D4A}
\definecolor{frameorange}{HTML}{B86A00}
\definecolor{fragmentgray}{HTML}{F7F7F8}
\definecolor{fragmentframe}{HTML}{8A8F98}

\tcbset{
  enhanced,
  breakable,
  boxrule=0.7pt,
  arc=1.8mm,
  left=2mm,right=2mm,top=1.5mm,bottom=1.5mm,
  coltitle=white,
  fonttitle=\bfseries,
}

\newtcolorbox{stageonebox}{%
  colback=promptblue,
  colframe=frameblue,
  title={Stage 1 Prompt: Reflection},
}
\newtcolorbox{stagetwobox}{%
  colback=promptgreen,
  colframe=framegreen,
  title={Stage 2 Prompt: Belief Update},
}
\newtcolorbox{stagethreebox}{%
  colback=promptorange,
  colframe=frameorange,
  title={Stage 3 Prompt: Decision},
}
\newtcolorbox{fragmentbox}[1]{%
  colback=fragmentgray,
  colframe=fragmentframe,
  title={#1},
}

\newcommand{\promptfont}{\ttfamily\small}
\newcommand{\ph}[1]{\textit{\{#1\}}}

\section{Ground Truth Model Construction}
\label{app:armote}

\subsection{The ARMOTE-MO Workflow}
\label{app:armote_algorithm}

\begin{algorithm}[H]
\caption{ARMOTE-MO: automated multi-output regression with multi-objective
         hyperparameter optimization}
\label{alg:armote}
\begin{algorithmic}[1]
\Require Composition--property dataset $\mathcal{D} = \{(\mathbf{x}_i, \mathbf{y}_i)\}_{i=1}^{N}$,
         search space $\Theta$, trial budget $T = 100$
\State Split $\mathcal{D}$ into $\mathcal{D}_\mathrm{train}$ / $\mathcal{D}_\mathrm{test}$ (80/20)
\State Fit standardization scalers for inputs and outputs on $\mathcal{D}_\mathrm{train}$;
       apply them to $\mathcal{D}_\mathrm{train}$ and $\mathcal{D}_\mathrm{test}$
\For{$t = 1$ to $T$}
    \State Sample $\theta_t \in \Theta$ with the Tree-Parzen Estimator
    \State Evaluate $\theta_t$ by 5-fold cross-validation on the standardized
           $\mathcal{D}_\mathrm{train}$, recording the mean MSE (standardized targets)
           and the mean $R^2$, each averaged across outputs
\EndFor
\State $\Theta^{*} \gets$ non-dominated set of $\{\theta_t\}$ in (MSE, $R^2$) space
       (minimize MSE, maximize $R^2$)
\State $\theta^{\dagger} \gets \arg\max_{\theta \in \Theta^{*}}$ cross-validated mean $R^2$
       (averaged across outputs)
\State Refit the multi-output RFR on all of the standardized $\mathcal{D}_\mathrm{train}$ at $\theta^{\dagger}$
\State \Return model, fitted scalers, and held-out $R^2$ on $\mathcal{D}_\mathrm{test}$
       (per output and averaged across outputs, in original target units)
\end{algorithmic}
\end{algorithm}

\subsection{Hyperparameter Search Spaces}
\label{app:armote_spaces}

\begin{table}[H]
\centering
\begin{tabular}{llc}
\toprule
Hyperparameter & Search range & Scale \\
\midrule
$n_\mathrm{trees}$           & [10, 150] & integer, uniform \\
\texttt{max\_depth}          & [10, 150] & integer, uniform \\
\texttt{min\_samples\_split} & [2, 20]   & integer, uniform \\
\texttt{min\_samples\_leaf}  & [1, 20]   & integer, uniform \\
\bottomrule
\end{tabular}
\caption{Hyperparameter search spaces explored by ARMOTE-MO for both alloy
         systems. Ranges are inclusive. All other random forest settings,
         including \texttt{max\_features}, were left at their scikit-learn
         defaults. The selected configurations are reported in
         Table~\ref{tab:rfr-hyperparameters}.}
\label{tab:armote-search-space}
\end{table}

\subsection{Truth-Function Generated Datasets}
\label{app:truth_datasets}

\section{Prompts Used by the LLM Allocation Agents}
\label{app:prompts}

\noindent The two LLM-based allocation strategies used separate two-call prompt
pipelines. Dynamic quantities inserted by the campaign code are shown in braces.
All prompt calls used GPT-4o through LangChain. The RHEA campaigns used
temperature $T = 0.2$, and the Fe-Co-Ni magnet campaigns used temperature
$T = 0.3$.

\subsection{Adaptive-\texorpdfstring{$\beta$}{β} Agent Prompts}
\label{app:prompts_adaptive_beta}

\begin{fragmentbox}{Adaptive-$\beta$ Call 1: $\beta$ selection}
\promptfont
You are selecting a beta parameter for a Bayesian Optimization campaign.
\vspace{0.35em}
Problem: \ph{problem\_description}\\
Objectives: \ph{objectives\_str}
\vspace{0.35em}
qEHVI focuses on what the GP currently believes is near-Pareto-optimal. It is the
reliable Pareto front improver and the right default in most situations.
\vspace{0.35em}
qUCB with elevated $\beta$ scouts design-space regions the GP has not characterized
well --- regions with high posterior uncertainty. Its purpose is to find productive
territory qEHVI has not yet been directed toward. Exploration will often produce no
short-term HV gain. That is acceptable when it locates new productive regions faster
than qEHVI alone would. It is NOT acceptable when exploration is saturated (MI
declining) or when the campaign is nearly complete.
\vspace{0.35em}
$\beta$ controls how far into uncertain regions qUCB will probe, measured in
posterior standard deviations. GP inputs and outputs are normalized, so $\beta$ has
the same meaning regardless of problem scale or physical units.
\vspace{0.35em}
\hspace*{1em}$\beta \approx 2$: Conservative. qUCB candidates stay near regions the GP already knows well. Right default when HV is actively improving.\\
\hspace*{1em}$\beta = 3$--$5$: Moderate exploration. Noticeably more uncertainty-seeking. Appropriate when HV progress has slowed but MI signals remaining territory.\\
\hspace*{1em}$\beta = 5$--$8$: Aggressive scouting. Sends candidates into poorly-characterized regions. Justified when HV has materially stalled and MI is still elevated.\\
\hspace*{1em}$\beta > 8$: Only in extreme stagnation with a strong MI signal.
\vspace{0.35em}
Early in the campaign (first \~{}20--25\% of resources consumed), the GP posterior is
poorly constrained --- uncertainty estimates are unreliable and elevated beta may send
candidates to arbitrary regions rather than genuinely informative ones. Prefer
$\beta$ near 2 unless you observe a specific signal that justifies deviation.
\vspace{0.35em}
$\beta$ only has effect if the selected option also allocates exploration points.
Choosing high $\beta$ with 0 explore points in the batch is a no-op.
\vspace{0.35em}
=== CAMPAIGN STATE ===\\
Iteration: \ph{state.iteration}\\
Resources consumed: \ph{budget\_pct}\%  |  Time remaining: \ph{time\_pct\_str}  |  Iters remaining: \~{}\ph{iters\_remaining}\\
Pareto front HV: \ph{hv\_line}\\
Pareto size: \ph{pareto\_size} points (\ph{pareto\_change})
\vspace{0.35em}
=== OPTIMIZATION SIGNALS ===\\
HV velocity ratio: \ph{vel\_str}\\
\hspace*{1em}(recent mean $\Delta$HV divided by historical baseline --- 1.0 = on pace, <0.5 = materially slowed, >1.0 = accelerating)\\
MI of best explore option: \ph{mi\_line} (\ph{mi\_vs\_str})\\
\hspace*{1em}(declining MI = design space increasingly well-characterized; elevated MI = unexplored territory remains)\\
\ph{pareto\_coverage}\\
=== RECENT HISTORY ===\\
\ph{recent\_history}
\vspace{0.35em}
=== OUTPUT FORMAT ===\\
Output EXACTLY these two lines (nothing else):\\
BETA: <number>\\
REASONING: <one sentence citing specific signals above>
\end{fragmentbox}

\begin{fragmentbox}{Adaptive-$\beta$ Call 2: allocation option selection}
\promptfont
You are selecting a resource allocation option for a Bayesian Optimization campaign.
\vspace{0.35em}
=== CONTEXT ===\\
Beta selected: \ph{beta:.1f}\\
Your reasoning: \ph{call1\_reasoning}\\
Resources consumed: \ph{budget\_pct}\%  |  Time remaining: \ph{time\_pct\_str}\\
Iters remaining: \~{}\ph{iters\_remaining}\\
Pareto front: \ph{hv\_pts\_str} pts, HV=\ph{hv\_hv\_str}
\vspace{0.35em}
=== HOW TO READ THE OPTIONS TABLE ===\\
Each option allocates the batch between qEHVI (exploitation) and qUCB (exploration) points.\\
qEHVI score: expected direct Pareto HV improvement --- compare qEHVI values across options only.\\
MI score: expected information gain about unexplored regions --- compare MI values across options only.\\
Do NOT compare qEHVI against MI --- they are on different scales.\\
Higher-index options send more points to qUCB at $\beta$=\ph{beta:.1f}.\\
Options with more qUCB points trade qEHVI improvement for MI gain.
\vspace{0.35em}
\ph{options\_table}
\vspace{0.35em}
=== OUTPUT FORMAT ===\\
Output EXACTLY these two lines (nothing else):\\
SELECTED\_OPTION: <integer 0 to \ph{num\_options - 1}>\\
REASONING: <1-2 sentences on why this option's qEHVI/MI tradeoff fits your beta reasoning>
\end{fragmentbox}

\subsection{Rule-Swap Agent Prompts}
\label{app:prompts_rule_swap}

\begin{fragmentbox}{Rule-Swap Call 1: assessment and $\beta$ selection}
\promptfont
You are guiding a Bayesian Optimization campaign.\\
Problem: \ph{problem\_description}\\
Objectives: \ph{objectives\_str}
\vspace{0.35em}
qEHVI already balances exploration and exploitation within the regions it believes the
Pareto front is likely to expand into. It naturally pushes toward unexplored areas
when those areas have high improvement potential. It is the right default in most
situations.
\vspace{0.35em}
qUCB serves a different purpose: it explores regions of the design space that have not
been evaluated much, regardless of whether the GP currently predicts them to be
Pareto-improving. Its value is in discovering areas the model has not seen yet ---
regions that may contain Pareto-optimal points qEHVI has never been directed toward.
\vspace{0.35em}
$\beta$ controls how explorative qUCB is: larger $\beta$ weights selection toward
points with greater GP uncertainty, pushing further into unmapped territory. Default
$\beta = 2.0$.
\vspace{0.35em}
MutualInfo measures how much the batch reduces GP uncertainty about the optimum. The
absolute value is not interpretable --- only compare it across options within the same
iteration. A higher MutualInfo option queries regions the GP understands least,
regions qEHVI tends to skip because it follows the surrogate's existing beliefs. It
does not directly guarantee better Pareto results, but it signals where the GP's
knowledge is thinnest and where qEHVI is most likely to be under-sampling.
\vspace{0.35em}
=== ITERATION HISTORY (last \ph{n\_completed} completed) ===\\
\ph{history\_text}
\vspace{0.35em}
=== RESOURCES ===\\
Iterations completed: \ph{state.iteration}\\
Estimated iters left: \ph{iters\_left}\\
Budget remaining: \$\ph{state.budget\_remaining:.2f} (\$\ph{state.cost\_per\_point:.2f}/point)\\
Time remaining: \ph{state.time\_remaining:.1f} weeks\\
Current Pareto front: \ph{pareto\_summary}
\vspace{0.35em}
As fewer iterations remain, decisions should shift from probing uncertain regions
toward acting on what the GP already knows is good: \~{}20+ iterations left allows
low-cost exploration, \~{}10 iterations left should prioritize high-confidence
improvements, and \~{}5 or fewer should exploit near-exclusively. If an event reduced
remaining iterations sharply, use the new iteration count.
\vspace{0.35em}
=== YOUR TASK ===\\
Examine the HV/pt column in the history above, but read it carefully in context.
\vspace{0.35em}
STEP 1 --- Identify which iterations were deliberate exploration. Rows where $\beta >
2.0$ or qUCB points > 0 represent intentional exploration investments. Low or zero
HV/pt on those rows is EXPECTED and NORMAL; do not count these rows as evidence of
crowding or stagnation.
\vspace{0.35em}
STEP 2 --- Assess the trend on qEHVI-primary iterations only. Crowding means HV/pt is
declining while Pareto points are still being added; stagnation means zero new Pareto
points across multiple consecutive qEHVI-primary iterations; healthy expansion means
HV/pt is stable or rising.
\vspace{0.35em}
STEP 3 --- Check whether past exploration paid off with a one-iteration lag. If the
most recent iteration used high $\beta$ or qUCB, do NOT raise $\beta$ further yet; give
qEHVI one iteration to exploit the updated model unless the row after exploration also
shows continued stagnation or crowding.
\vspace{0.35em}
STEP 4 --- Factor in iterations remaining. With very few iterations left, raising
$\beta$ is riskier. When in doubt with few iters left, prefer $\beta = 2.0$.
\vspace{0.35em}
Respond using EXACTLY this format:\\
ASSESSMENT: <1-2 sentences describing what the HV/pt trend shows --- crowding, stagnation, or healthy expansion>\\
BETA: <number 1--100; default is 2.0>\\
REASONING: <1 sentence explaining the beta choice>
\end{fragmentbox}

\begin{fragmentbox}{Rule-Swap Call 2: allocation option selection}
\promptfont
You are selecting a resource allocation for a Bayesian Optimization campaign.
\vspace{0.35em}
=== SITUATION ASSESSMENT ===\\
\ph{assessment}
\vspace{0.35em}
=== AVAILABLE OPTIONS ===\\
\ph{options\_text}
\vspace{0.35em}
Note: qEHVI score and MutualInfo are on different scales --- compare each metric
across options separately; do not compare them to each other. The MutualInfo absolute
value is not meaningful on its own: only use it to rank options against each other. A
notably higher MutualInfo on explore-heavy options signals those options query regions
the GP knows least --- use this alongside the crowding/stagnation assessment when
deciding how many qUCB points to include.
\vspace{0.35em}
=== DECISION RULES ===\\
Option 0 (all qEHVI) is the DEFAULT. Choose it unless the assessment above identifies
crowding or stagnation that warrants qUCB exploration.\\
\hspace*{1em}$\bullet$ Crowding detected: choose an option with some qUCB points; scale the number of qUCB points to the severity of the signal.\\
\hspace*{1em}$\bullet$ Stagnation detected: use a larger qUCB share.\\
\hspace*{1em}$\bullet$ Estimated iters left is very low (\~{}5 or fewer): prefer Option 0 regardless of crowding.\\
\hspace*{1em}$\bullet$ No clear signal, or early in the campaign: Option 0.
\vspace{0.35em}
Resources: est. iters left = \ph{iters\_left} | budget \$\ph{state.budget\_remaining:.2f}
\vspace{0.35em}
Respond using EXACTLY this format:\\
SELECTED\_OPTION: <integer 0 to \ph{num\_options - 1}>\\
REASONING: <1-2 sentences connecting the assessment to this specific choice>
\end{fragmentbox}

\subsection{Event Handling in Agent Prompts}
\label{app:prompts_events}

Resource events enter the agent prompts through the campaign-state and remaining-iterations fields above. The Rule-Swap prompt additionally instructs the model to treat an event-shortened campaign according to the new estimated iteration count rather than the original budget.

\section{Supplementary Results}
\label{app:supplementary_results}

\subsection{Per-Experiment Cumulative Mutual Information}
\label{app:cumulative_mi_table}

\begin{table}[H]
\centering
\small
\begin{tabular}{lllccc}
\toprule
System & Experiment & Strategy & Cumulative MI (nats) & Final HV & AUC-HV \\
\midrule
RHEA & Baseline & Adaptive-$\beta$ & $19.54 \pm 0.41$ & $0.9967 \pm 0.00019$ & $51.59 \pm 0.092$ \\
 &  & Rule-Swap & $17.20 \pm 0.50$ & $0.9970 \pm 0.00018$ & $51.79 \pm 0.097$ \\
 &  & Fixed Mixed & $15.48 \pm 0.34$ & $0.9971 \pm 0.00017$ & $50.98 \pm 0.093$ \\
 &  & qEHVI & $14.23 \pm 0.45$ & $0.9983 \pm 0.00018$ & $52.02 \pm 0.077$ \\
\midrule
RHEA & Budget cut & Adaptive-$\beta$ & $12.29 \pm 0.38$ & $0.9843 \pm 0.00071$ & $21.87 \pm 0.078$ \\
 &  & Rule-Swap & $10.93 \pm 0.43$ & $0.9861 \pm 0.00048$ & $21.99 \pm 0.086$ \\
 &  & Fixed Mixed & $11.48 \pm 0.42$ & $0.9855 \pm 0.00043$ & $21.26 \pm 0.070$ \\
 &  & qEHVI & $8.78 \pm 0.31$ & $0.9869 \pm 0.00063$ & $22.11 \pm 0.084$ \\
\midrule
RHEA & Time cut & Adaptive-$\beta$ & $11.97 \pm 0.37$ & $0.9848 \pm 0.00054$ & $21.84 \pm 0.081$ \\
 &  & Rule-Swap & $10.53 \pm 0.35$ & $0.9858 \pm 0.00044$ & $21.95 \pm 0.062$ \\
 &  & Fixed Mixed & $10.63 \pm 0.28$ & $0.9846 \pm 0.00054$ & $21.30 \pm 0.093$ \\
 &  & qEHVI & $8.86 \pm 0.39$ & $0.9866 \pm 0.00100$ & $22.00 \pm 0.089$ \\
\midrule
RHEA & Dual & Adaptive-$\beta$ & $6.11 \pm 0.35$ & $0.9341 \pm 0.00602$ & $7.33 \pm 0.073$ \\
 &  & Rule-Swap & $5.25 \pm 0.26$ & $0.9413 \pm 0.00423$ & $7.37 \pm 0.073$ \\
 &  & Fixed Mixed & $5.32 \pm 0.18$ & $0.9488 \pm 0.00216$ & $7.10 \pm 0.072$ \\
 &  & qEHVI & $4.66 \pm 0.33$ & $0.9486 \pm 0.00318$ & $7.43 \pm 0.075$ \\
\midrule
Fe-Co-Ni & Baseline & Adaptive-$\beta$ & $19.63 \pm 0.48$ & $0.8539 \pm 0.00976$ & $38.23 \pm 0.317$ \\
 &  & Rule-Swap & $16.74 \pm 0.47$ & $0.8457 \pm 0.00871$ & $38.26 \pm 0.254$ \\
 &  & Fixed Mixed & $15.34 \pm 0.48$ & $0.8555 \pm 0.00963$ & $38.09 \pm 0.288$ \\
 &  & qEHVI & $13.17 \pm 0.34$ & $0.8438 \pm 0.00932$ & $38.37 \pm 0.306$ \\
\midrule
Fe-Co-Ni & Budget cut & Adaptive-$\beta$ & $14.60 \pm 0.50$ & $0.8186 \pm 0.00880$ & $18.26 \pm 0.122$ \\
 &  & Rule-Swap & $11.99 \pm 0.47$ & $0.8057 \pm 0.00900$ & $18.20 \pm 0.203$ \\
 &  & Fixed Mixed & $10.82 \pm 0.40$ & $0.7992 \pm 0.00627$ & $18.05 \pm 0.132$ \\
 &  & qEHVI & $9.67 \pm 0.37$ & $0.8155 \pm 0.00850$ & $18.46 \pm 0.180$ \\
\midrule
Fe-Co-Ni & Time cut & Adaptive-$\beta$ & $14.66 \pm 0.50$ & $0.8018 \pm 0.00618$ & $17.37 \pm 0.095$ \\
 &  & Rule-Swap & $12.45 \pm 0.58$ & $0.8178 \pm 0.00874$ & $17.62 \pm 0.123$ \\
 &  & Fixed Mixed & $11.43 \pm 0.62$ & $0.7941 \pm 0.00547$ & $17.17 \pm 0.156$ \\
 &  & qEHVI & $10.44 \pm 0.48$ & $0.7999 \pm 0.00721$ & $17.44 \pm 0.131$ \\
\midrule
Fe-Co-Ni & Dual & Adaptive-$\beta$ & $14.69 \pm 0.50$ & $0.8033 \pm 0.00751$ & $18.11 \pm 0.136$ \\
 &  & Rule-Swap & $12.28 \pm 0.34$ & $0.8111 \pm 0.00854$ & $18.41 \pm 0.163$ \\
 &  & Fixed Mixed & $10.83 \pm 0.42$ & $0.8046 \pm 0.00697$ & $18.08 \pm 0.185$ \\
 &  & qEHVI & $9.97 \pm 0.39$ & $0.8056 \pm 0.00697$ & $18.39 \pm 0.143$ \\
\bottomrule
\end{tabular}

\caption{Cumulative mutual information, final normalized hypervolume, and area
         under the hypervolume curve for every system, experiment, and strategy
         (mean $\pm$ SEM, $n = 25$ seeds).}
\label{tab:cumulative-mi-full}
\end{table}

\subsection{Pairwise Significance Tests}
\label{app:significance}

All strategy comparisons use the two-sided Wilcoxon signed-rank test applied to
per-seed differences. The two strategies in a pair are matched by shared random seed,
so each of the $n = 25$ paired observations contrasts strategies that began from the
same initial design and the same random state. No multiple-comparison correction is
applied: the 144 tests reported in Table~\ref{tab:significance} (six strategy pairs
$\times$ three metrics $\times$ eight experiments) are individually uncorrected, and
$p$-values close to the conventional $0.05$ threshold should therefore be read as
descriptive rather than confirmatory.

\begin{table}[H]
\centering
\footnotesize
\begin{tabular}{lllccc}
\toprule
System & Experiment & Strategy pair & $p$ (final HV) & $p$ (AUC-HV) & $p$ (cum.\ MI) \\
\midrule
RHEA & Baseline & Adaptive-$\beta$ vs.\ Rule-Swap & $0.090$ & $0.006$\textsuperscript{**} & $<0.001$\textsuperscript{***} \\
 &  & Adaptive-$\beta$ vs.\ Fixed Mixed & $0.080$ & $<0.001$\textsuperscript{***} & $<0.001$\textsuperscript{***} \\
 &  & Adaptive-$\beta$ vs.\ qEHVI & $<0.001$\textsuperscript{***} & $<0.001$\textsuperscript{***} & $<0.001$\textsuperscript{***} \\
 &  & Rule-Swap vs.\ Fixed Mixed & $0.958$ & $<0.001$\textsuperscript{***} & $0.006$\textsuperscript{**} \\
 &  & Rule-Swap vs.\ qEHVI & $<0.001$\textsuperscript{***} & $<0.001$\textsuperscript{***} & $<0.001$\textsuperscript{***} \\
 &  & qEHVI vs.\ Fixed Mixed & $<0.001$\textsuperscript{***} & $<0.001$\textsuperscript{***} & $0.026$\textsuperscript{*} \\
\midrule
RHEA & Budget cut & Adaptive-$\beta$ vs.\ Rule-Swap & $0.059$ & $0.210$ & $0.011$\textsuperscript{*} \\
 &  & Adaptive-$\beta$ vs.\ Fixed Mixed & $0.075$ & $<0.001$\textsuperscript{***} & $0.182$ \\
 &  & Adaptive-$\beta$ vs.\ qEHVI & $0.007$\textsuperscript{**} & $0.017$\textsuperscript{*} & $<0.001$\textsuperscript{***} \\
 &  & Rule-Swap vs.\ Fixed Mixed & $0.275$ & $<0.001$\textsuperscript{***} & $0.210$ \\
 &  & Rule-Swap vs.\ qEHVI & $0.096$ & $0.001$\textsuperscript{**} & $<0.001$\textsuperscript{***} \\
 &  & qEHVI vs.\ Fixed Mixed & $0.012$\textsuperscript{*} & $<0.001$\textsuperscript{***} & $<0.001$\textsuperscript{***} \\
\midrule
RHEA & Time cut & Adaptive-$\beta$ vs.\ Rule-Swap & $0.173$ & $0.071$ & $0.017$\textsuperscript{*} \\
 &  & Adaptive-$\beta$ vs.\ Fixed Mixed & $0.937$ & $<0.001$\textsuperscript{***} & $0.007$\textsuperscript{**} \\
 &  & Adaptive-$\beta$ vs.\ qEHVI & $0.005$\textsuperscript{**} & $0.007$\textsuperscript{**} & $<0.001$\textsuperscript{***} \\
 &  & Rule-Swap vs.\ Fixed Mixed & $0.200$ & $<0.001$\textsuperscript{***} & $0.874$ \\
 &  & Rule-Swap vs.\ qEHVI & $0.059$ & $0.020$\textsuperscript{*} & $0.001$\textsuperscript{**} \\
 &  & qEHVI vs.\ Fixed Mixed & $0.001$\textsuperscript{**} & $<0.001$\textsuperscript{***} & $<0.001$\textsuperscript{***} \\
\midrule
RHEA & Dual & Adaptive-$\beta$ vs.\ Rule-Swap & $0.191$ & $0.692$ & $0.012$\textsuperscript{*} \\
 &  & Adaptive-$\beta$ vs.\ Fixed Mixed & $0.042$\textsuperscript{*} & $0.001$\textsuperscript{**} & $0.080$ \\
 &  & Adaptive-$\beta$ vs.\ qEHVI & $0.011$\textsuperscript{*} & $0.148$ & $<0.001$\textsuperscript{***} \\
 &  & Rule-Swap vs.\ Fixed Mixed & $0.230$ & $<0.001$\textsuperscript{***} & $0.596$ \\
 &  & Rule-Swap vs.\ qEHVI & $0.059$ & $0.002$\textsuperscript{**} & $0.032$\textsuperscript{*} \\
 &  & qEHVI vs.\ Fixed Mixed & $0.937$ & $<0.001$\textsuperscript{***} & $0.027$\textsuperscript{*} \\
\midrule
Fe-Co-Ni & Baseline & Adaptive-$\beta$ vs.\ Rule-Swap & $0.596$ & $0.833$ & $<0.001$\textsuperscript{***} \\
 &  & Adaptive-$\beta$ vs.\ Fixed Mixed & $0.634$ & $0.958$ & $<0.001$\textsuperscript{***} \\
 &  & Adaptive-$\beta$ vs.\ qEHVI & $0.672$ & $0.411$ & $<0.001$\textsuperscript{***} \\
 &  & Rule-Swap vs.\ Fixed Mixed & $0.711$ & $0.731$ & $0.010$\textsuperscript{**} \\
 &  & Rule-Swap vs.\ qEHVI & $0.653$ & $0.542$ & $<0.001$\textsuperscript{***} \\
 &  & qEHVI vs.\ Fixed Mixed & $0.542$ & $0.542$ & $<0.001$\textsuperscript{***} \\
\midrule
Fe-Co-Ni & Budget cut & Adaptive-$\beta$ vs.\ Rule-Swap & $0.312$ & $0.596$ & $<0.001$\textsuperscript{***} \\
 &  & Adaptive-$\beta$ vs.\ Fixed Mixed & $0.045$\textsuperscript{*} & $0.080$ & $<0.001$\textsuperscript{***} \\
 &  & Adaptive-$\beta$ vs.\ qEHVI & $0.711$ & $0.560$ & $<0.001$\textsuperscript{***} \\
 &  & Rule-Swap vs.\ Fixed Mixed & $0.791$ & $0.560$ & $0.071$ \\
 &  & Rule-Swap vs.\ qEHVI & $0.367$ & $0.191$ & $<0.001$\textsuperscript{***} \\
 &  & qEHVI vs.\ Fixed Mixed & $0.312$ & $0.026$\textsuperscript{*} & $0.037$\textsuperscript{*} \\
\midrule
Fe-Co-Ni & Time cut & Adaptive-$\beta$ vs.\ Rule-Swap & $0.411$ & $0.127$ & $0.011$\textsuperscript{*} \\
 &  & Adaptive-$\beta$ vs.\ Fixed Mixed & $0.474$ & $0.141$ & $<0.001$\textsuperscript{***} \\
 &  & Adaptive-$\beta$ vs.\ qEHVI & $0.711$ & $0.812$ & $<0.001$\textsuperscript{***} \\
 &  & Rule-Swap vs.\ Fixed Mixed & $0.080$ & $0.027$\textsuperscript{*} & $0.127$ \\
 &  & Rule-Swap vs.\ qEHVI & $0.164$ & $0.075$ & $0.001$\textsuperscript{**} \\
 &  & qEHVI vs.\ Fixed Mixed & $0.833$ & $0.141$ & $0.164$ \\
\midrule
Fe-Co-Ni & Dual & Adaptive-$\beta$ vs.\ Rule-Swap & $0.525$ & $0.164$ & $<0.001$\textsuperscript{***} \\
 &  & Adaptive-$\beta$ vs.\ Fixed Mixed & $0.615$ & $0.791$ & $<0.001$\textsuperscript{***} \\
 &  & Adaptive-$\beta$ vs.\ qEHVI & $0.474$ & $0.063$ & $<0.001$\textsuperscript{***} \\
 &  & Rule-Swap vs.\ Fixed Mixed & $0.367$ & $0.067$ & $0.011$\textsuperscript{*} \\
 &  & Rule-Swap vs.\ qEHVI & $0.711$ & $0.958$ & $<0.001$\textsuperscript{***} \\
 &  & qEHVI vs.\ Fixed Mixed & $0.833$ & $0.134$ & $0.113$ \\
\bottomrule
\end{tabular}

\caption{Pairwise strategy comparisons for final normalized hypervolume, AUC-HV,
         and cumulative mutual information ($n = 25$ seeds per strategy). Two-sided
         paired Wilcoxon signed-rank, uncorrected; markers denote
         \textsuperscript{*}$p < 0.05$, \textsuperscript{**}$p < 0.01$,
         \textsuperscript{***}$p < 0.001$.}
\label{tab:significance}
\end{table}

\end{document}